\documentclass[reprint,amsmath,amssymb,aps,pre,longbibliography,showkeys]{revtex4-1}
\usepackage{hyperref}
\usepackage{graphics}
\usepackage[FIGTOPCAP]{subfigure}
\usepackage{bm}
\usepackage{tikz}
\usepackage{natbib}

\usetikzlibrary{shapes,arrows}

\newcommand{\grad}{\vec{\nabla}}
\newcommand{\Dif}[2]{\frac{\partial #1}{\partial #2}}
\newcommand{\dif}[1]{\partial_{#1}}
\newcommand{\ave}[1]{\langle #1 \rangle}
\definecolor{hlcolor}{rgb}{1,0.2,0}
\definecolor{ntcolor}{rgb}{0,0.6,0}

\begin{document}

\title{Vortex with four-fold defect lines in a simple model of self-propelled particles}

\keywords{self-propelled particle, confinement, defect line, suppressed spreading}

\date{\today}

\author{Hamid Seyed-Allaei}
\affiliation{Department of Physics, Sharif University of Technology, P. O. Box 11155-9161, Tehran, Iran}
\author{Mohammad Reza Ejtehadi}
\email{ejtehadi@sharif.edu}
\affiliation{Department of Physics, Sharif University of Technology, P. O. Box 11155-9161, Tehran, Iran}

\begin{abstract}
We studied formation of vortex with four-fold symmetry in a minimal model of self-propelled particles, confined inside a squared box, using computer simulations and also theoretical analysis. In addition to the vortex pattern, we observed five other phases in the system: homogeneous gaseous phase, band structures, moving clumps, moving clusters and vibrating rings. All six phases emerge from controlling strength of noise and contribution of repulsion and alignment interactions. We studied shape of the vortex and its symmetry in detail. The pattern shows exponential defect lines where incoming and outgoing flows of particles collide. We show that alignment and repulsion interactions between particles are necessary to form such patterns. Finally, we derived hydro-dynamical equations for our model and compared them with the results of both computer simulations and Quincke rotors. A good agreement between the three is observed.
\end{abstract}

\maketitle

\section{Introduction}

All of us have seen fascinating movement patterns of flocks of birds~\cite{Biro2006,Nagy2010,Ballerini2008} or schools of fishes~\cite{Becco2006,Makris2006}. Similar structures are widely seen in many places, ranging from our body size~\cite{Helbing2000a} down to nano meter scales~\cite{Surrey1997,Schaller2010}, including either living individuals~\cite{Czirok1996,Peruani2012} or non living ones~\cite{Narayan2007,Baker2010}. The common feature between all these diverse systems is activity among the individuals and therefore these systems are called active matter. Active matters, because of consumption and injection of energy, are always out of equilibrium and in recent years have attracted many attentions\cite{Vicsek2012,Marchetti}. In one of the primary efforts, collective patterns from basic local interactions has been produced~\cite{Vicsek1995}. Later, phenomenological theories as well as microscopic descriptions were established to find the characteristics and features of active matter~\cite{Toner1995,Toner1998,Baskaran2008a,Tailleur2008}. In addition, more complex collective patterns were observed by introduction of new models~\cite{Farrell2012,Shimoyama1996,McCandlish2012}, and helped to improve knowledge of the phase transitions and behavior of active matter~\cite{Chate2008a,romanczuk2011brownian,romanczuk2012mean,Solon2015,Caussin}.

Vortices are one of the most interesting patterns observed in active matter. Swirling of daphnia around a light shaft~\cite{Ordemann2003}, rotation of bacteria in droplets of bacillus subtilis morphotype which grow on an agar substrate~\cite{Czirok1996} and vortices that are formed by the moving actin filaments on a surface coated with heavy meromyosin~\cite{Schaller2010} are biological examples of vortex formation. There are also non-biological samples of vortex formation, e.g. vertically vibrated granular rods~\cite{Blair2008}, anisotropic rods in a container~\cite{Kudrolli2008} and micrometer sized insulator spheres well known as Quincke rotors~\cite{Bricard2013}. In addition to stable vortex patterns, dissipation of vortices like a turbulent phase is observed in colonies of bacteria~\cite{Cisneros2011,Dunkel2013}. Many studies have been done to understand vortex formation, its nature and characteristics. For example a model of self-propelled particles that repel each other in close distances and attract in far makes a giant vortex~\cite{DOrsogna2006}. Another example is the case of particles with intrinsic curvature in their motion like microtubules moving on a surface coated with myosines~\cite{Sumino2012}. More sophisticated models consider chemotaxis and proliferation for bacteria to understand their swirling~\cite{Czirok1996}. It is also possible to have a vortex array in the system~\cite{Großmann2014,Nagai2015}, acquiring self-propelled particles with alignment and anti-alignment interactions with respect to distance~\cite{Großmann2014} or time correlated noise~\cite{Nagai2015}.

In the experimental study of Quincke rotors~\cite{Bricard2013} the particles and their interactions produce a complex vortex inside a squared box. The vortex shows a four-fold symmetry. This four-fold shape is along with an effect that we call hereafter ``suppressed spreading''. In suppressed spreading, particles which are bounced back from a corner tend to spread over all available directions, but because of the flow of the other particles and collisions, the spreading of the outgoing flow is suppressed and is limited to a smaller angle. Suppressed spreading is visible as a curved boundary, where the direction and density of particles change spontaneously, and we call it ``defect line''.

Inspired by the experimental vortex formations~\cite{Surrey1997,Bricard2013}, here we study vortex pattern of self-propelled particles, confined in a geometry. Such a study has been done by simulation of active granular particles with inelastic interactions, confined in a squared box~\cite{Y.LimonDuparcmeur1995,Grossman2008}. However, these studies are limited in size and they could not describe the complex structures and density jump lines of the Quincke rotors observed in the experiments~\cite{Bricard2013}.

First we introduce a minimal model and find the key elements required to have a vortex with the symmetry of confining geometry (section \ref{section:model}). The model is a simple generalization of continuous Vicsek model~\cite{Czirok1996} with alignment and repulsion interactions. Both interactions have physical interpretation and are derived theoretically in reference~\cite{Bricard2013}. The important role of repulsion in vortex formation is revealed recently~\cite{Bricard2015}. Next, we derive continuum hydrodynamic equations in the limits of high and low noise to compare the solutions with the particle model and experimental results (section \ref{section:theory}). Finally we present simulation results and discuss about patterns and their characteristic (\ref{section:simulation}). 

\section{Model}
\label{section:model}

We consider two dimensional self-propelled particles with the same constant speed $v$. Angle of the velocity with $x$ axis is $\theta$ and the direction of motion for each particle is toward $\hat{e}_\theta$. The direction of each particle is changed by torque. This torque is originated from particle-particle and wall-particle interactions. For the dynamics of particles we consider
\begin{equation}
\dot{\vec{r}}_i = v \hat{e}_{\theta_i},
\label{eq:r-dynamics}
\end{equation}
\begin{equation}
\dot{\theta}_i = \tau^p_{i} + \tau^w_{i}  + \epsilon \eta_{i}(t),
\label{eq:theta-dynamics-1}
\end{equation}
where $\vec{r}_i$ is the position of the $i$th particle and $\hat{e}_{\theta_i}$ is a unit vector along swimming direction of the particle with angel $\theta_i$, ($\hat{e}_{\theta} = \cos(\theta_i) \hat{e}_x + \sin(\theta_i) \hat{e}_y$). In Eq.~(\ref{eq:theta-dynamics-1}), $\tau_i^p$ and $\tau^w_i$ are the torques acting on the particle $i$ from the other particles and the walls, respectively. We added a noise term $\epsilon \eta_{i}(t)$ which represents stochastic behavior of self-propelled particles and their environment, where $\eta_i(t)$ is a Gaussian uncorrelated white noise with $\ave{\eta_i(t)} = 0$ and $\ave{\eta_i(t) \eta_i(t^\prime)}=\delta(t-t^\prime)$, where $\epsilon$ is the noise amplitude.

\begin{figure}
\centering
\includegraphics[width=\columnwidth,clip]{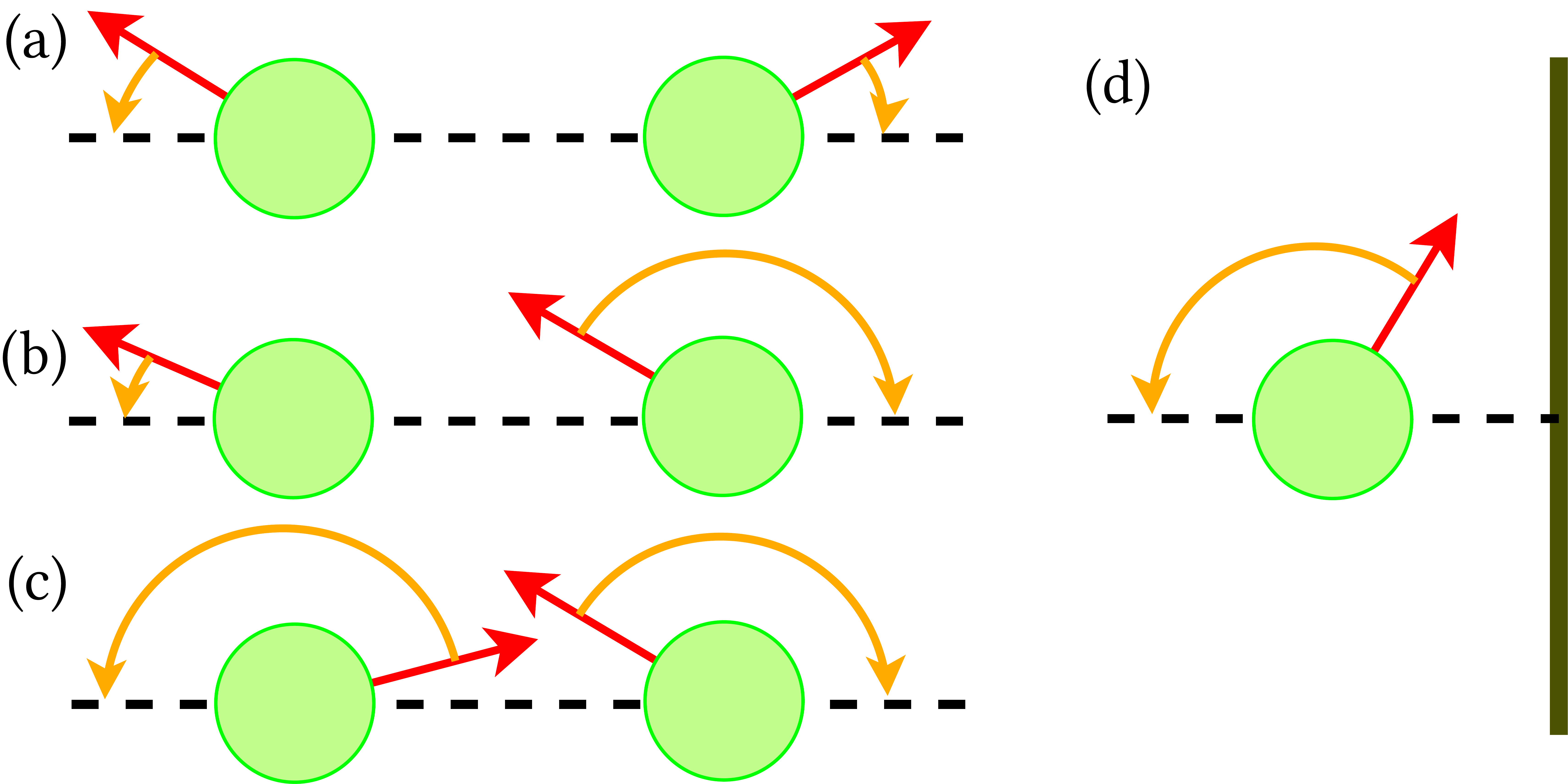}
\caption{(Color online) Repulsive torque between two particles in Eq.~(\ref{eq:torque-particle}) (a, b, c), and between a particle and a wall in Eq.~(\ref{eq:torque-wall}) (d). Each particle is represented by a green disk. The straight red arrow shows direction of the particle. The arced orange arrow shows the direction of rotation, end of this arrow is particle's final direction and its length has no relation to the magnitude of the torque. The thick vertical line is a wall. The dashed line in (a,b,c) is the inter distance line between two colliding particles and in (d) is the perpendicular line to the wall. In (a) particles are escaping and the torque intensifies this escape. In (b) particles are moving together and the torques cause them to move away. In (c) particles are moving toward one another and the torque inhibits them reaching each other. In (d) particle is moving toward the wall.}
\label{fig:Interaction}
\end{figure}

The particle-particle interaction is a combination of alignment and repulsion. Alignment means particles rotate to make their moving directions parallel to each other and repulsion means that particles rotate to run away from each other (Fig.~\ref{fig:Interaction}). The alignment and repulsion interactions that are close to Quincke Rotors interactions~\cite{Bricard2013} give,
\begin{equation}
\begin{aligned}
\tau_{ji} &= \frac{g_p}{\pi} A(r_{ji}) \times \\
&\left[ (1 - \alpha) sin \left( \theta_j - \theta_i \right) + \alpha (\hat{r}_{ji}\times\hat{e}_{\theta_i}) \cdot \hat{e}_z \right].
\label{eq:torque-particle}
\end{aligned}
\end{equation}
The first term on the right hand side is aligning and the second term is repulsive torque. $\vec{r}_{ji}$ is the distance vector from particle $j$ to $i$, $g_p$ is the strength of particle-particle interaction and $A(r_{ji})$ is a function of interparticle distance. $0 \leq \alpha \leq 1$ controls the relative contribution of repulsion and alignment terms, i.e. $\alpha=0$ corresponds to the original continuous Vicsek model, and $\alpha=1$ corresponds to a fully repulsive particle system. It is also interesting to see the result of negative $\alpha$ that is a combination of alignment and attraction. The $\times$ sign between vectors shows vector product and $.$ shows dot product. The model is two dimensional, but for a compact presentation we use dot product of z direction unit vector, $\hat{e}_z$ with the result of vector product as a scalar value.

Figure \ref{fig:Interaction} shows a schematic presentation of interaction terms. As one can see, repulsion turns velocities of particles in opposite of their inter distance direction.

For simplicity here we restrict ourselves to the case of constant $\alpha$ and consider a Heaviside step function for $A$
\begin{equation}
A(r_{ji}) = \Theta(R - r_{ji}),
\label{eq:Q}
\end{equation}
where $R$ indicates the range of interaction between the particles.

Very similar to the repulsive interaction between particles, if we label a wall by $w$, the applied torque by the wall $w$ on a particle $i$ is given by:
\begin{equation}
\tau_{wi} = \frac{g_w}{\pi r_{wi}} A(r_{wi}) \left[ (\hat{n}_w \times\hat{e}_{\theta_i}) \cdot \hat{e}_z \right],
\label{eq:torque-wall}
\end{equation}
where $\hat{n}_{w}$ is the unit normal vector of the wall, $r_{wi}$ is the distance of the particle from the wall, and $g_w$ is the strength of the particle-wall interaction. The factor $1 / r_{wi}$ guaranties that the particles never pass the wall.

One should note that in contrast to the alignment torque, the repulsion torque will not conserve the total angular momentum. In models with spontaneous vortex formation, a generating source of angular momentum is necessary.

In the next section we will use this microscopic model to obtain hydrodynamic equations of the system.

\section{Hydrodynamic equations}
\label{section:theory}

We can characterize the state of the system with density of particles, $\rho$, and their polarization vector $\vec{\mathcal{P}} = \ave{\hat{e}_{\theta}}$ in space. The governing hydrodynamic equations of density and polarization could be derived from microscopic equations. With the method presented in reference~\cite{Dean1996}, we derive Fokker-Plank equation for particle density up to second order of spatial derivatives,
\begin{equation}
\begin{aligned}
&\frac{d f (\theta,\vec{r})}{dt} = - (1 - \alpha) g_p \dif{\theta} \\
&\Bigg[ f (\theta,\vec{r}) \int_0^{2 \pi} d\theta^{\prime} \sin(\theta^{\prime} - \theta) \\
&\big( R^2 f(\theta^{\prime},\vec{r}) + \frac{R^4}{8} \nabla^2 f(\theta^{\prime}, \vec{r}) \big)  \Bigg] \\
& -\dif{\theta} \left[ \frac{\alpha g_p R^3}{3} f(\theta,\vec{r}) \vec{\nabla} \rho(\vec{r}) \cdot (\hat{e}_{\theta} \times \hat{e}_z) \right] \\
&+ D_r \frac{\partial^2 f (\theta,\vec{r})}{\partial \theta^2} - v \hat{e}_{\theta} \cdot \grad{f (\theta,\vec{r})},
\end{aligned}
\label{eq:fokker-plank}
\end{equation}
where $f (\theta,\vec{r})$ is the density of particles at point $\vec{r}$ moving in direction $\theta$, and $D_r = \epsilon^2 / 2$ is rotational diffusion of particles. The terms on the right hand side represent alignment, repulsion, diffusion and advection of the particles, respectively.

In the following we solve the Fokker-Plank equation for two different regimes of high noise and low noise.

\subsection{High Noise Limit}
\label{subsection:high-noise-limit}

We use Fourier transform of density function $f(\theta,\vec{r}) = \frac{1}{2\pi} \sum_k e^{\imath k\theta} \tilde{f}_k$, where $\imath$ is the imaginary number and $\tilde{f}_k$ is the $k$th Fourier component of $f$. Thanks to linear independency of Fourier basis $e^{\imath k \theta}$, we can split Eq.~(\ref{eq:fokker-plank}) to an infinite set of separate recurrence equations in Fourier space:
\begin{equation}
\begin{aligned}
&\frac{d \tilde{f}_k (\vec{r})}{dt} = - D_r k^2 \tilde{f}_k + \frac{(1 - \alpha) g_p R^2 k}{2} \times \\
&\Big( \tilde{f}_{k-1} (\tilde{f}_1 + \frac{R^2}{8} \nabla^2 \tilde{f}_1) \\
&- \tilde{f}_{k+1} (\tilde{f}_{-1} + \frac{R^2}{8} \nabla^2 \tilde{f}_{-1}) \Big) \\
&- \frac{\alpha g_p R^3 k}{6} \vec{\nabla} \rho \cdot \Big( \hat{x} (\tilde{f}_{k-1} - \tilde{f}_{k+1} ) \\
&- \imath \hat{y} (\tilde{f}_{k-1} + \tilde{f}_{k+1} ) \Big) \\
&- v \partial_x \frac{\tilde{f}_{k-1} + \tilde{f}_{k+1}}{2} - v \partial_y \frac{\tilde{f}_{k-1} - \tilde{f}_{k+1}}{2\imath} .
\end{aligned}
\label{eq:fourier-fokker-plank}
\end{equation}

Because the set of equations is unlimited we need to truncate it at some point. There is a damping term $-D_r k^2 \tilde{f}_k$ with time scale $\tau_k = -D_r k^2$, meaning that higher moments of $\tilde{f}_k$ vanish faster. Here we assume that moments of $\tilde{f}_k$ for $k \geq 3$ are zero and the second moments converge to their equilibrium values fast enough that we can assume $\dot{\tilde{f}}_{\pm 2} = 0$. This assumption is valid until the damping terms for higher moments are dominant. Comparing the coefficients of the right hand side of Eq.~(\ref{eq:fourier-fokker-plank}) with $\tau_k^{-1}$ when $k=3$, we will get conditions, $D_r \gg \frac{(1-\alpha) g_p R^2}{6}$, $D_r \gg \frac{\alpha g_p R^3}{18}$ and $D_r \gg \frac{v}{9}$ to truncate Eq.~(\ref{eq:fourier-fokker-plank}) for $k \geq 3$. Given in the values of $R=1$, $\alpha = 0.5$, $g_p = 2$ and $v=1$ (the same as simulations) we find inequality $D_r \gg 0.16$ which satisfies all conditions.

After the truncation, we can find $\tilde{f}_{\pm 2}$ in terms of $\tilde{f}_{\pm 1}$ and $\tilde{f}_0$, then one can replace it in the equations of $\dot{\tilde{f}}_{\pm 1}$.	 $\tilde{f}_0$ is density of particles and $\tilde{f}_{\pm 1}$ is related to the polarization of particles. Defining $\vec{W} \equiv \rho \vec{\mathcal{P}}$, we can write $\tilde{f}_{\pm 1} = W_x \mp \imath W_y$. Simplifying equations and solving the real and imaginary parts separately, it results in the continuity equation,
\begin{equation}
\Dif{\rho}{t} + v \grad \cdot \vec{W} = 0,
\label{eq:continuity}
\end{equation}
and evolution of $\vec{W}$,
\begin{equation}
\begin{aligned}
&\dot{\vec{W}} = \Bigg[ \frac{(1 - \alpha) \rho g_p R^2}{2} - D_r \\
&- \frac{(1 - \alpha)^2 g^2_p R^4 W^2}{8D_r} - \frac{\alpha^2 g^2_pR^6}{72 D_r} \left| \vec{\nabla} \rho \right|^2 \Bigg] \vec{W}\\
&+ \Bigg[ \left( \frac{v^2}{16 D_r} + \frac{(1 - \alpha) \rho g_p R^4}{16} \right) \nabla^2 \vec{W}\\
&- \frac{(1 - \alpha)^2 g^2_p R^6}{32D_r} \left( \vec{W} \cdot \nabla^2 \vec{W} \right) \vec{W} \Bigg]\\
&- \frac{(1 - \alpha) g_p v R^2}{16D_r} \big[ - \frac{5}{2} \grad W^2 \\
&+ 5 \vec{W} \grad \cdot \vec{W} + 3 \vec{W} \cdot \grad \vec{W} \big]\\
&- \left( \frac{\alpha g_p R^3}{6} \rho + \frac{v}{2} \right) \grad \rho \\
&+ \frac{\alpha (1 - \alpha) g^2_pR^5}{12 D_r} (\vec{W} \cdot \grad \rho) \vec{W}  \\
&+ \frac{\alpha g_p v R^3}{48 D_r} \Big( 2 \vec{W} \nabla^2 \rho + 2 \grad \rho \cdot \grad \vec{W} \\
&- 3 \grad \cdot \vec{W} \grad \rho + 3 \grad \rho \times (\grad \times \vec{W})  \Big).
\end{aligned}
\label{eq:hd-high-noise}
\end{equation}
On the right hand side of Eq.~(\ref{eq:hd-high-noise}), the first bracket contains driving terms which cause spontaneous polarization, with the first three terms that usually appear in active matter hydrodynamic equations~\cite{Toner1995,Toner2005,Baskaran2008,Bertin2006}. The fourth term introduces a reduction in polarity due to the net repulsive torque in density gradient. The second bracket is very important for spread of polarization. It is a diffusion-like term for $\vec{W}$ and comes from the alignment of neighboring particle, closer than distance $R$. The third bracket comes from alignment interaction between particles and is well known in both phenomenological and analytical studies~\cite{Toner1995,Toner2005,Baskaran2008,Bertin2006}. Rest of the equation shows escape of particles from higher densities due to repulsion and advection. We are not worried about $D_r^{-1}$ factor in the equation as we are in limit $D_r \gg 0.16$.

\begin{figure}
	\centering
	\subfigure[$\rho(\vec{r})$]{\label{fig:compare-continuum-density-strip}\includegraphics[width=0.49\columnwidth]{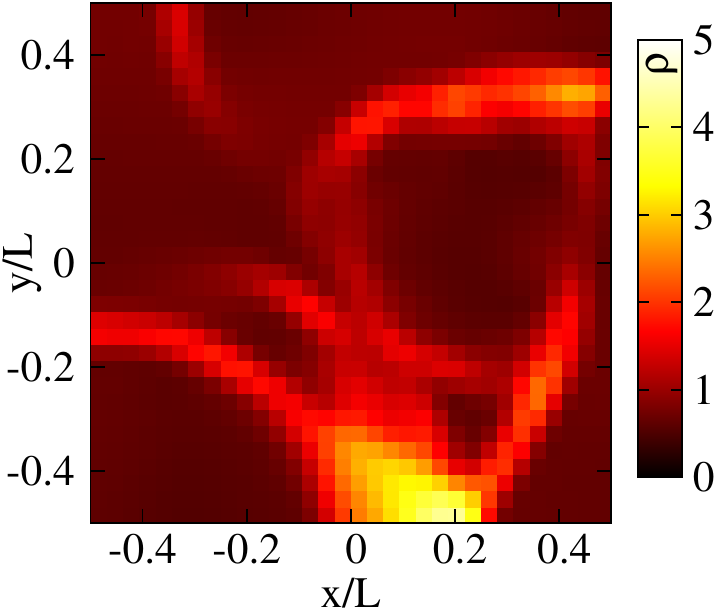}}%
	\subfigure[$\vec{v}(\vec{r})$]{\label{fig:compare-continuum-v-strip}\includegraphics[width=0.49\columnwidth]{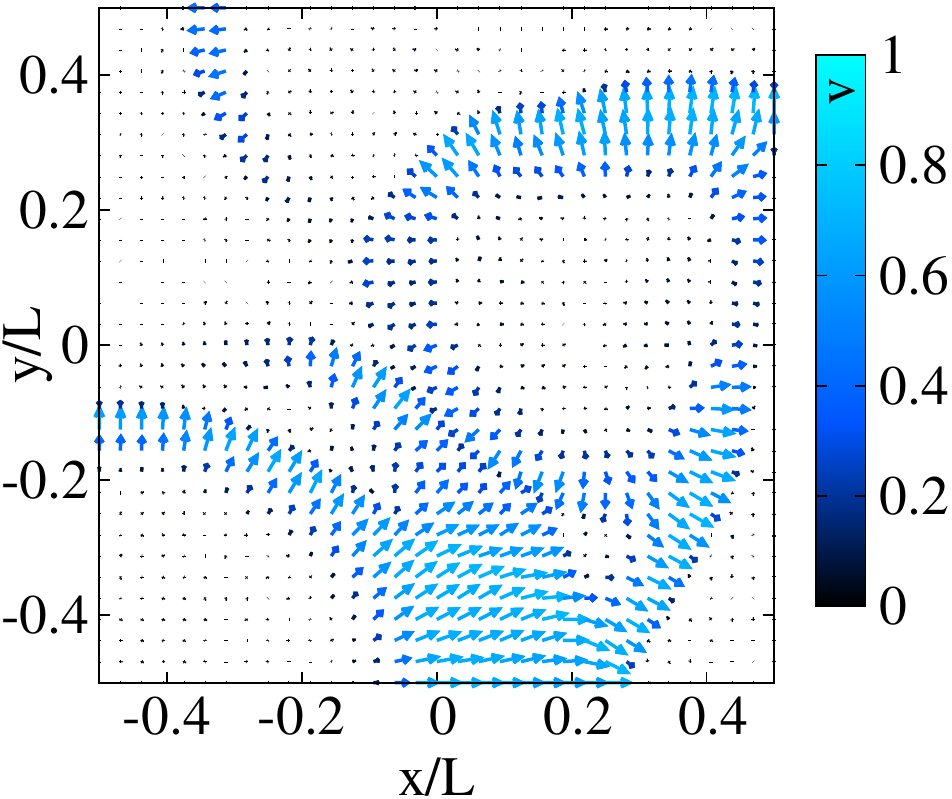}}
\caption{(Color online) Results of continuum model [Eq.~(\ref{eq:hd-high-noise})], before occurrence of divergence with initial density $\rho_0 = 1$, alignment interaction $g = 2$, repulsion factor $\alpha = 0.5$ and noise $D_r=0.59$ ($\epsilon = 1.08$) in a box with size of $120$. \subref{fig:compare-continuum-density-strip} Density profile. \subref{fig:compare-continuum-v-strip} Velocity field, the lengths of arrows is proportional to the magnitude. Both images correspond to the same moment and their grid size is 32 by 32.}
\label{fig:compare-strip}
\end{figure}

Looking for steady homogeneous solutions of Eqs. (\ref{eq:hd-high-noise}) and (\ref{eq:continuity}) with initial density $\rho_0$, one can find a critical noise value $\epsilon_c = R \sqrt{(1-\alpha)g_p \rho_0}$ ($D_C = \epsilon_c^2 / 2$), at which the system behavior changes from polar to non-polar homogeneous state. This change of behavior is also important for confined particles, because as simulations show, once the system goes to a polar state at this critical noise, particles start to rotate. With our simulation parameters one finds $D_C = 0.5$, therefore Eqs. (\ref{eq:hd-high-noise}) and (\ref{eq:continuity}) are not able to explain the system for low noise values far from the transition point ($D_r \gg 0.16$). Solving the equations numerically, the answer for $D_r \geq 0.6$ is non-polar and homogeneous ($\vec{W} = \vec{0}$) and with $D_r=0.59$ numerical instabilities emerge and no vortex is observed. But interestingly, we can observe stripes forming and propagating in the box and reflecting from corners at the initial stage of computation before divergence occurs (Fig.~\ref{fig:compare-strip}).

To obtain more stable hydrodynamic equations we derive them in the low noise limit. In the low noise regime, density and polarization change slowly in space, that  enhance the stability of equations.

\subsection{Low Noise Limit}

In this limit, we decompose $f(\theta,\vec{r})$ to the number density $\rho(\vec{r})$ and orientational probability distribution of particles, $P(\theta,\vec{r})$ [i.e. $f(\theta,\vec{r}) = P(\theta,\vec{r}) \rho(\vec{r})$]. We multiply both sides of Eq.~(\ref{eq:fokker-plank}) by an arbitrary function $b(\theta)$, and integrate over $\theta$ to find the governing dynamics of $\langle b(\theta) \rangle$. Then for $b(\theta)=1$ it gives us continuity equation in terms of $\ave{\hat{e}_{\theta}} = \vec{\mathcal{P}}$ and $\rho$ [$\dot{\rho} = - v \grad \cdot ( \rho(\vec{r}) \ave{\hat{e}_{\theta}})$]. Substituting $\dot{\rho}$ from continuity equation into the governing equation of $\langle b(\theta) \rangle$, we find,
\begin{equation}
\begin{aligned}
& \rho(\vec{r}) \frac{d \ave{b(\theta)}}{dt}  = (1-\alpha) g R^2 \\
&\times \ave{b^\prime(\theta) \sin(\theta^\prime - \theta)}_{\theta^\prime,\theta} \\
&\times \left( \rho(\vec{r})^2 + \frac{R^2}{8} \rho(\vec{r}) \nabla^2 \rho(\vec{r}) \right)  \\
&+ \frac{1}{8} (1-\alpha) g R^4 \rho(\vec{r})^2 \\
&\times \ave{b^\prime(\theta) \nabla^2 \ave{\sin(\theta^\prime - \theta)}_{\theta^\prime}}_\theta\\
&+ \frac{1}{4} (1-\alpha) g R^4 \rho(\vec{r}) \grad \rho(\vec{r}) \\
&\cdot \ave{b^\prime(\theta) \grad \ave{\sin(\theta^\prime - \theta)}_{\theta^\prime}}_\theta\\
&+ \frac{\alpha g_p R^3}{3} \rho(\vec{r}) \grad \rho(\vec{r}) \cdot (\ave{b^\prime(\theta) \hat{e}_{\theta}} \times \hat{e}_z) \\
&- v \grad \cdot ( \rho(\vec{r}) \ave{b(\theta) \hat{e}_{\theta}}) \\
&+ v \ave{b(\theta)} \grad \cdot ( \rho(\vec{r}) \ave{\hat{e}_{\theta}})\\
&+ D_r \rho(\vec{r}) \ave{b^{\prime\prime}(\theta)} + O(\nabla^3),
\end{aligned}
\label{eq:b-evolution}
\end{equation}
where $\ave{}_\theta$ indicates average over $\theta$ with probability distribution $P(\theta,\vec{r})$. In the low noise limit, $P(\theta,\vec{r})$ is sharply peaked around the mean and we can approximate $\sin(\theta^\prime - \theta)$ with $\theta^\prime - \theta$ to find homogeneous solutions of Eq.~(\ref{eq:b-evolution}) for $b(\theta)=\theta$ and $b(\theta)=\theta^2$. This gives us dynamical equation for variance of $\theta$,
\begin{equation}
\frac{d \sigma^2_\theta}{dt} = D_r - (1-\alpha) g_p \rho_0 \sigma_\theta^2 ,
\end{equation}
where $D_r$ is acting here as a source of dispersion for $\theta$, while alignment interaction $(1-\alpha) g_p \rho$ reduces the dispersion. From this equation we find that $\sigma^2_\theta$ decays with time scale $\tau_{\sigma} = 1 / (1-\alpha) g_p \rho$ to its equilibrium value $\sigma^2_\theta = D_r / (1-\alpha) g_p \rho$. Our assumption for sharpness of distribution function is true when $\sigma^2_\theta \ll 1$ and this gives us a limit for the noise, $D_r \ll (1-\alpha) g_p \rho$. By expanding $e^{\imath \theta}$ up to second order around the mean value $\theta_m$ and using the facts that $\ave{e^{\pm \imath \theta}} = \mathcal{P}_x \pm \imath \mathcal{P}_y$ and $e^{\pm \imath \theta_m} = \frac{1}{\mathcal{P}} ( \mathcal{P}_x +\pm\imath \mathcal{P}_y)$, we can find a relation between $\sigma^2_\theta$ and polarization $\mathcal{P}$ as, $\mathcal{P} = (1 - \sigma^2_\theta/2)$. The same assumption - small deviation - in Eq.~(\ref{eq:b-evolution}) with $b(\theta) = e^{\imath \theta}$ helps to find the dynamics of the polarization, $\vec{\mathcal{P}}(\vec{r},t)$,

\begin{equation}
\begin{aligned}
& \Dif{\vec{\mathcal{P}}}{t}  = \vec{\mathcal{P}} \Big[ 2 (1-\alpha) g R^2 \\
&\times ( \rho(\vec{r}) + \frac{R^2}{8} \nabla^2 \rho(\vec{r}) ) (1 - \mathcal{P}) - D_r \Big] \\
&+ \frac{1}{8} (1-\alpha) g R^4 \rho(\vec{r}) \Big[(2\mathcal{P} - 1) \nabla^2 \vec{\mathcal{P}}\\
&- (4 \mathcal{P} - 3) \hat{\mathcal{P}} \hat{\mathcal{P}} \cdot \nabla^2 \vec{\mathcal{P}} \Big] \\
&+ \frac{1}{4} (1-\alpha) g R^4 \Big[(2\mathcal{P}-1) \grad \rho(\vec{r}) \cdot \grad \vec{\mathcal{P}}\\
&- \frac{4\mathcal{P} - 3}{2 \mathcal{P}^2} \vec{\mathcal{P}} \grad \rho(\vec{r}) \cdot \grad \mathcal{P}^2 \Big] \\
&+ \frac{\alpha g_p R^3}{3} \Big[ (1-2P) \grad \rho \\
&+ (4P-3) \hat{\mathcal{P}} \cdot \grad \rho \hat{\mathcal{P}} \Big] \\
&+ v \Big[ \frac{2}{\rho} \grad \left( \rho (\mathcal{P}-1) \right) \\
&+ \frac{1}{\rho} \hat{\mathcal{P}} \cdot \grad \left( \rho (\mathcal{P}-1)(\mathcal{P}-3) \hat{\mathcal{P}} \right)\\
&+ (\mathcal{P}-1)(\mathcal{P}-3) \hat{\mathcal{P}} \grad \cdot \hat{\mathcal{P}} - \vec{\mathcal{P}} \cdot \grad \vec{\mathcal{P}} \Big] \\
&+ O(\nabla^3).
\end{aligned}
\label{eq:polarization-evolution}
\end{equation}
The first bracket on the right hand side of Eq.~(\ref{eq:polarization-evolution}) prevents $\mathcal{P}$ to become zero, the second bracket spreads polarization because of interaction with neighboring particles, the third bracket shows an alignment competition between high and low density regions, the fourth bracket represents repulsion of particles (moving against $\grad \rho$) and the last one shows advection.

It is clear that solving Eq.~(\ref{eq:polarization-evolution}) together with continuity equation in squared geometry is complicated, therefore we restrict our calculation to circular box where the scalar variables depend only on radial component $r$. We also neglect second order derivations that are related to shorter changes of fields. Even with these assumptions, presence of a non-homogeneous steady state is not obvious and we need to write everything up to the first order of $D_r / (1-\alpha)g_p \rho$. With all these simplifications one finds that continuity equation implies that polarization is toward polar coordinate unit vector $\hat{\mathcal{\phi}}$, and the density satisfies the following relation:
\begin{equation}
\rho(r) + \frac{D_r}{(1-\alpha)g_p} \ln(\rho) = \frac{3 v}{\alpha g} \ln \left( \frac{r}{r_0} \right),
\label{eq:density-r}
\end{equation}
where $r_0$ is a length scale related to the initial density of particles. We can see that at distances shorter than $r_0$ the density gets very small values. Similar results with logarithmic dependence of density is obtained in Quincke rotors~\cite{Bricard2015}. If the system has no noise ($D_r=0$) the right hand side of Eq.~(\ref{eq:density-r}) is negative for $r < r_0$ that corresponds to a zero density in center. In fact no particle reaches the center in noiseless system, but when turning on the noise some particles could reach the center with a small chance. These results are in good agreement with simulation results (see Figs. \ref{fig:radial-density} and \ref{fig:compare-density-dist} in section \ref{section:simulation}).

Integrating equation (\ref{eq:density-r}) for $D_r=0$ over the box and equating it with the total number of particles, one obtains an expression for $r_0$,
\begin{equation}
\rho_0 = \frac{3 v}{2 \alpha g_p} \left[ \frac{r_0^2}{R_{\mathrm{box}}^2} - 1 - 2 \ln \left( \frac{r_0}{R_{\mathrm{box}}} \right) \right],
\label{eq:r_0}
\end{equation}
where $R_{box}$ is the radius of the circular box. Eq.~(\ref{eq:r_0}) shows $r_0$ decreases either by increasing $\rho_0$ or $\alpha g_p$, both causing stronger steric repulsions. Although our dynamics is not Newtonian, we see dependence of $r_0/R_{box}$ on $v$. Having a mixture of particles with different velocities, this effect can separate fast and slow particles. We will discuss this effect and phase separation of slow and fast particles in a future work. It is interesting to mention that a mixture of fast and slow particles could be prepared experimentally by using different sizes of Quincke rotors due to their velocity relation,
\begin{equation}
v(E_0,a) = a C \sqrt{(E_0/E_Q)^2 - 1},
\label{eq:Quicke-rotor-velocity}
\end{equation}
where $v$ is the velocity of particles, $a$ is their radius, $C$ is a combination of elements of modified mobility matrix and Maxwell-Wagner time of the environment, $E_0$ is the applied electric field, $E_Q$ is a threshold for electric field. In this relation $E_Q$ and $C$ depend on both environment and $a$~\cite{Bricard2013}.

\subsection{Numerical Method in Computation of Noiseless System}
\label{subsection:numerical-method}

To find steady state solution of Eq.~(\ref{eq:polarization-evolution}), when $D_r=0$ ($\mathcal{P}=1$), we need to rewrite Eq.~(\ref{eq:polarization-evolution}) in terms of $\vec{W}$ to avoid numerical instabilities (subsection \ref{subsection:high-noise-limit}),

\begin{equation}
\begin{aligned}
\dot{\vec{W}} &= \frac{1}{8} (1-\alpha) g_p R^4 \\
&\times \left( \rho (\nabla^2 \vec{W} - \vec{p} \cdot \nabla^2 \vec{W} \vec{p}) \right) \\
&- \frac{\alpha g_p R^3}{3} \rho(\vec{r}) (\hat{e}_z \times \vec{p}) (\hat{e}_z \times \vec{p}) \cdot \grad \rho(\vec{r}) \\
&- v \left( \vec{p} \grad \cdot \vec{W} + \vec{W} \cdot \grad \vec{p} \right) + O(\nabla^3).
\end{aligned}
\label{eq:noiseless-W-dynamic}
\end{equation}

To compare theoretical result with simulations and experiment, we also numerically integrate Eqs. (\ref{eq:noiseless-W-dynamic}) and (\ref{eq:continuity}) with a square box geometry. Integration of the equations are done using pseudo spectral method, semi-implicit time stepping and anti-aliasing (2/3 rule) techniques~\cite{uecker2009short,canuto1993spectral}. Using these techniques alone, is not enough to have stability and we need to add a $K \nabla^2 \rho$ ($K = 0.2$) term to the continuity equation to avoid negative density values. This extra term does not drastically change the underlying physics of the model and this technique was used before in the literature~\cite{Farrell2012}. Additionally at the initial stage of integration we must set $v=0.005$ and slowly increase the speed up to the highest stable value which is $v = 0.3$. We set $\alpha=0.5$ and $g_p=2$ to compare our result with the particle model. A circular rotating pattern with homogeneous density for initial condition is necessary for stability. With the use of image method we apply slip boundary condition. That means we replicate our system on the sides with reflecting $\vec{W}$, to justify slip boundary condition.

Now we have the right theoretical tools to compare our simulation results. In the next section we will provide the method and the results of simulations based on the microscopic model to compare with the theory.

\section{Simulation}
\label{section:simulation}

\begin{figure}
\centering
	\includegraphics[width=\columnwidth]{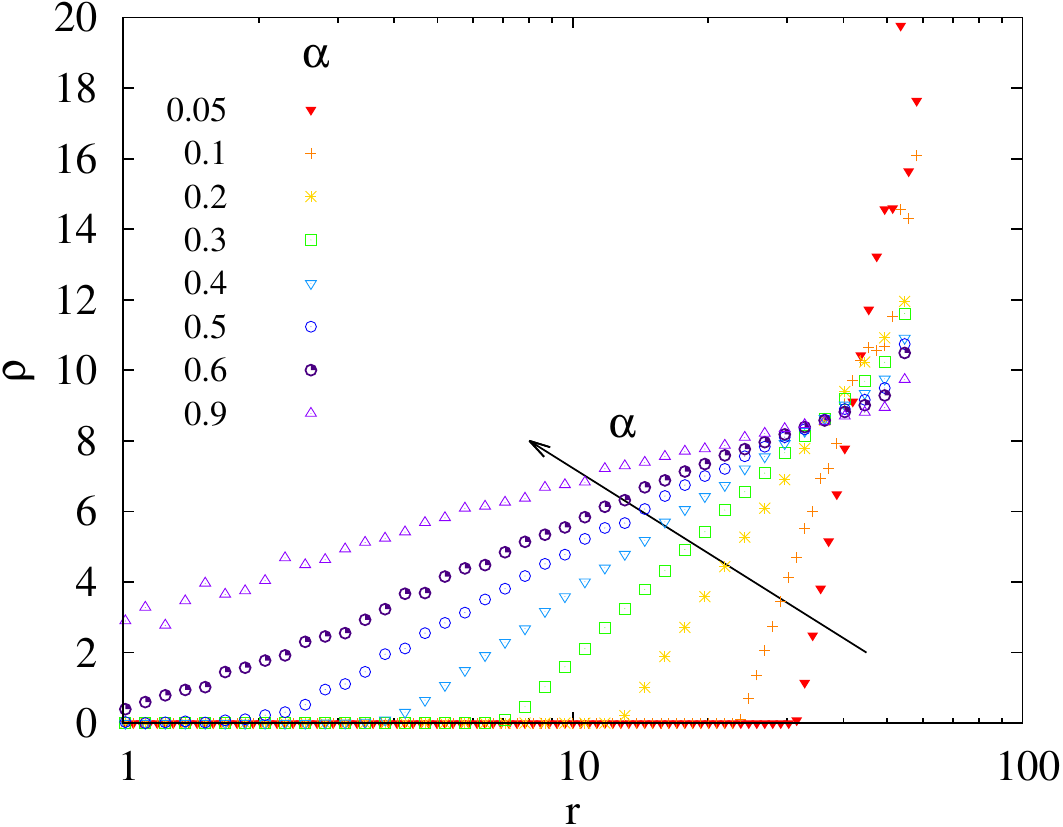}
\caption{(Color online) Density of particles $\rho(r)$ as a function of radial distance $r$, for a noiseless system in a circular box with diameter $L=120$ and initial density $\rho_0 = 8$. Labels show different values of $\alpha$. One can see that the density increases logarithmically with $r$ [Eq.~(\ref{eq:density-r})].}
\label{fig:radial-density}
\end{figure}

To simulate our model, we used integration technique (Ito method~\cite{longtin2010stochastic}) with time steps $dt=5\times10^{-4}$. We set $R = 1$, $g_p = 2$, $g_w = 40$. The speed of particles is set equal to one ($v = 1$), unless otherwise stated. The control parameters in simulations are the strength of noise $\epsilon$, repulsion strength $\alpha$, and initial density $\rho_0$. Except for part \ref{subsection:circularbox} where we use a circular box, we always put $N$ particles inside a squared box with four surrounding walls of size $L$. Simulation box must be large enough to capture all aspects of a self-propelled system. This limit is originated from band structures~\cite{Gregoire2004,Chate2008,Bertin2009} that have typical length scale of $v / \epsilon_c^2$. Then for given parameters, one finds the condition $L \gg v / \epsilon_c \sim 4$. We initially position particles homogeneously in a triangular lattice with uniformly random direction of motions. We run the systems long enough to guarantee that the effects of initial condition vanish, and then sample the systems.

\subsection{Circular Box}
\label{subsection:circularbox}

\begin{figure}
\centering
	\subfigure[]{\label{fig:compare-density-dist:slope}\includegraphics[width=\columnwidth,clip]{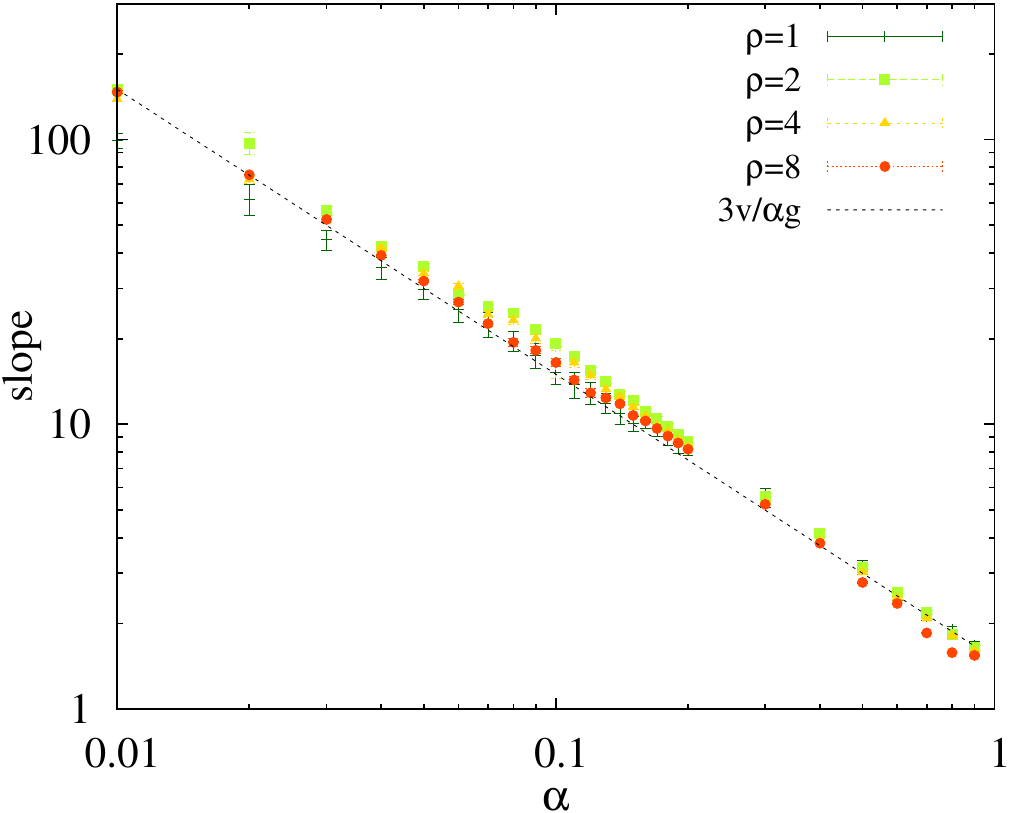}}
	\subfigure[]{\label{fig:compare-density-dist:r_0}\includegraphics[width=\columnwidth,clip]{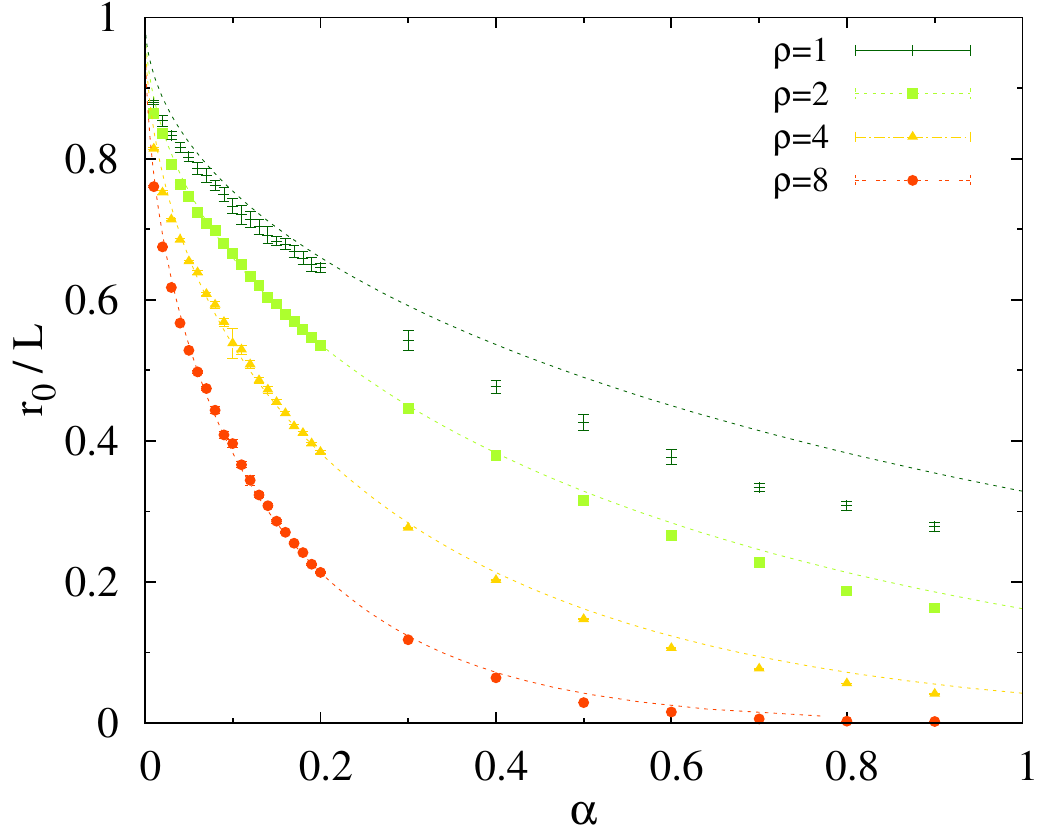}}
\caption{(Color online) Comparison of simulations (points with error bars) with theoretical Eqs. (\ref{eq:density-r}) and (\ref{eq:r_0}) (dashed lines) of a circular box with radius $R_{box}=60$ and different initial densities. \subref{fig:compare-density-dist:slope} The plot shows slope in Eq.~(\ref{eq:density-r}) as a function of $\alpha$. \subref{fig:compare-density-dist:r_0} The plot shows $r_0$ as a function of $\alpha$ with different $\rho_0$ [Eq. (\ref{eq:r_0})].}
\label{fig:compare-density-dist}
\end{figure}

First we use a circular boundary to compare with the solutions of our theoretical Eqs. (\ref{eq:density-r}) and (\ref{eq:r_0}). Figure \ref{fig:radial-density} shows particle density as a function of radial coordinate, $r$, for noiseless microscopic model in a circular box with radius $R_{box}=60$, and initial density $\rho_0=8$. The figure shows the results for different values of $\alpha$. Density is increasing logarithmically with $r$, and by increasing $\alpha$, reduction in the slopes and the size of empty region in center is visible. Equation (\ref{eq:density-r}) predicts exactly the same behavior. For a more precise comparison we also plot the slopes and $r_0$ values versus $\alpha$ for given density values $\rho=1,2,4,8$ in Fig.~\ref{fig:compare-density-dist}. Dashed lines in Fig.~\ref{fig:compare-density-dist} show Eqs. (\ref{eq:density-r}) and (\ref{eq:r_0}) for each given density. We observe a good agreement between theory and simulation specially for higher densities. This is because in high density our assumption of having sharp distribution for $\theta$ is more accurate.

\subsection{Phases}

\begin{figure}
	\centering
	\subfigure[$\epsilon = 1.2$]{\label{snapshots:homogeneous}\includegraphics[width=0.45\columnwidth]{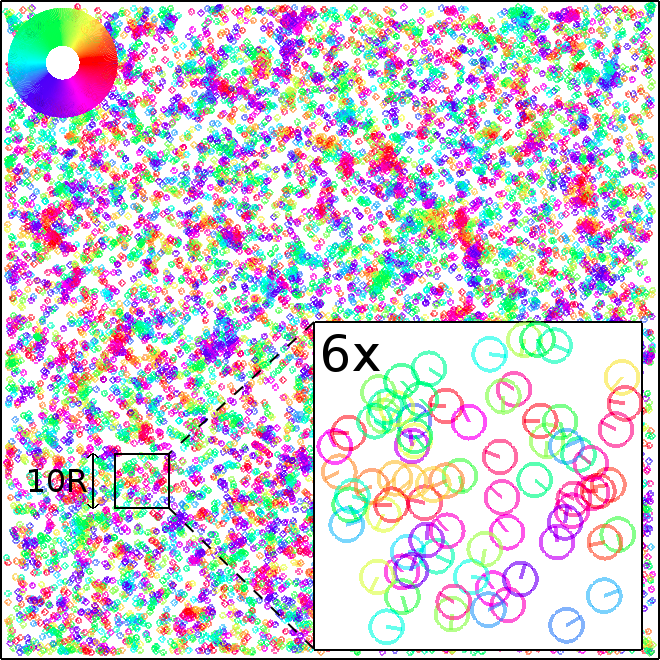}}
	\subfigure[$\epsilon = 0.8$]{\label{snapshots:strip}\includegraphics[width=0.45\columnwidth]{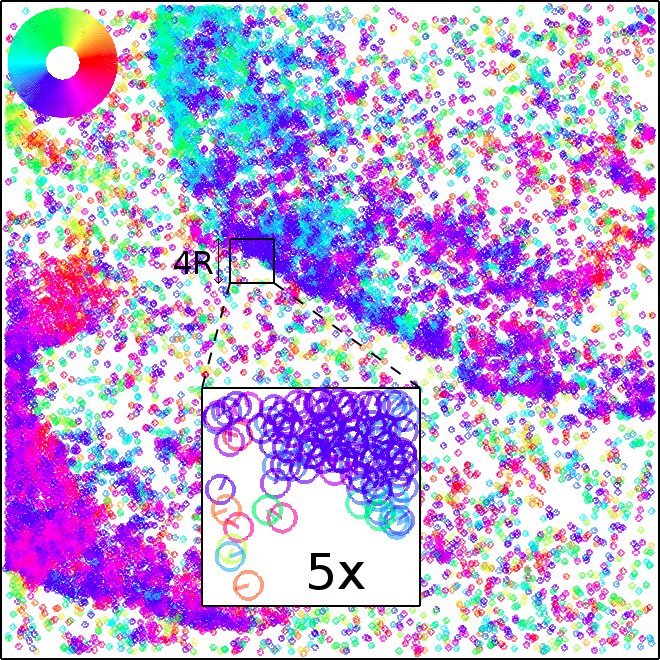}}
	\subfigure[$\epsilon = 0$]{\label{snapshots:vortex}\includegraphics[width=0.45\columnwidth]{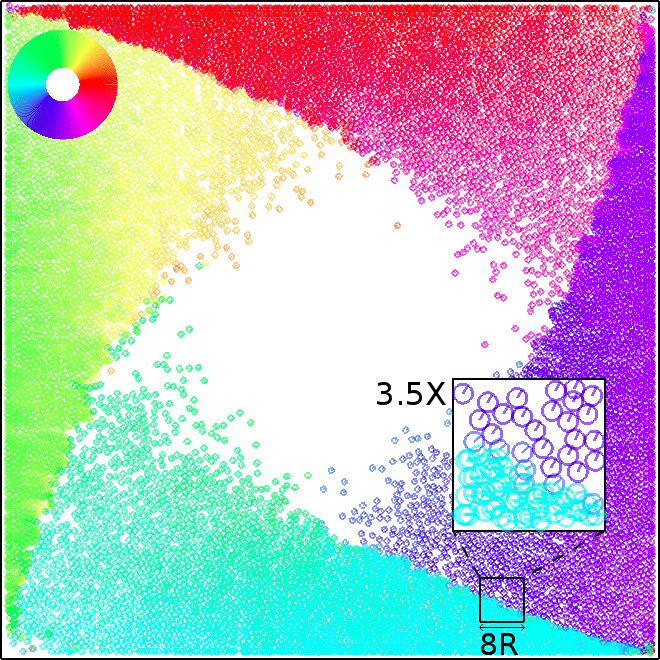}}
\caption{(Color online) Snapshots of the simulations with various noises and fixed repulsion factor $\alpha = 0.5$, initial density $\rho_0=1$ and box size of $L=120$. Open circles indicate particles, color shows the direction of motion corresponding to the color wheel located at top left corner of each box, and particles tails are along their trajectories. Squared windows represent zooming in a part of the box, the length of selected area and the magnification factor is written close to the corresponding window. By decreasing the noise, one can see \subref{snapshots:homogeneous} homogeneous, \subref{snapshots:strip} inhomogeneous with curved stripes and \subref{snapshots:vortex} vortex. In the vortex phase, suppressed spreading and four defect lines are observable.}
\label{fig:snapshots}
\end{figure}

During simulations by changing control parameters $\epsilon$ and $\alpha$, we observed six different regimes (phases): homogeneous gaseous, band structures, moving clumps, moving clusters, vibrating rings and vortex. At a given value of $\alpha  = 0.5$, and by decreasing noise, homogeneous gaseous phase, band structures and vortex pattern are observed (see Fig.~\ref{fig:snapshots}). Starting from very high noise, the system is in a gaseous phase [Fig.~\ref{fig:snapshots}\subref{snapshots:homogeneous}]. By decreasing the noise strength, the density gets inhomogeneous and particles form traveling curved stripes which are due to the reflections from the walls and the corners [Fig.~\ref{fig:snapshots}\subref{snapshots:strip}]. By further reducing the noise, particles start to rotate in the box as shown in Fig.~\ref{fig:snapshots}\subref{snapshots:vortex}. The direction of rotation is random and depends implicitly on initial positions and the string of random number samples. In rotation [Fig.~\ref{fig:snapshots}\subref{snapshots:vortex}], one can easily recognize suppressed spreading and the presence of the defect lines. The same pattern has been experimentally observed in the suspension of Quincke rotors~\cite{Bricard2013}.

\begin{figure}
	\centering
	\subfigure[$\rho = 1$, $\alpha = 0.01$]{\label{snapshots:low_alpha}\includegraphics[width=0.45\columnwidth]{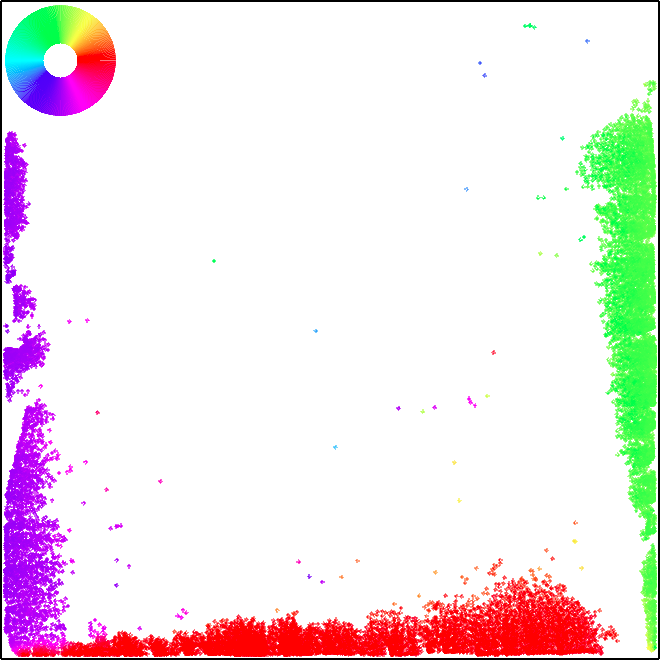}}
	\subfigure[$\rho = 1$, $\alpha = 0$]{\label{snapshots:zero_alpha}\includegraphics[width=0.45\columnwidth]{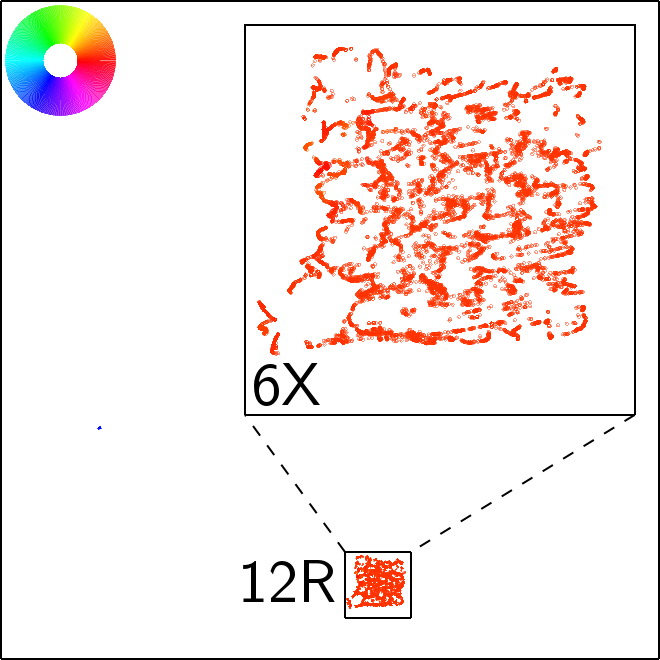}}
	\subfigure[$\rho = 1$, $\alpha = -0.01$]{\label{snapshots:negative_alpha}\includegraphics[width=0.45\columnwidth]{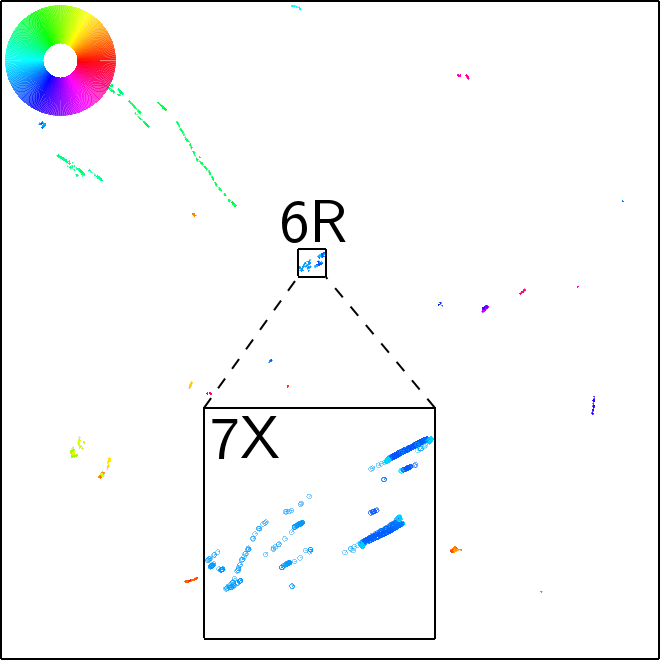}}
	\subfigure[$\rho = 1$, $\alpha = -0.2$]{\label{snapshots:strong_negative_alpha}\includegraphics[width=0.45\columnwidth]{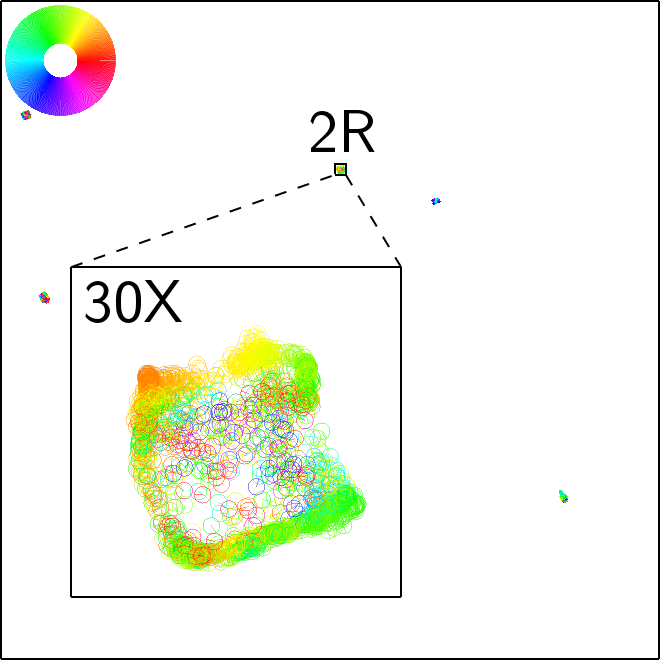}}
\caption{(Color online) Snapshots of the simulated system with $\rho_0=1$, $\epsilon = 0.1$ and different values of $\alpha$. Dots indicate particles, color shows the direction of motion corresponding to the color wheel located at top left corner of each box.  Squared windows represent zooming in a part of the box, the length of selected area and the magnification factor is written close to the corresponding window. \subref{snapshots:low_alpha} The system with a very small value of $\alpha$ where particles rotate cohesively around the box  and a very small defect line is observable. \subref{snapshots:zero_alpha} Switching off the repulsion ($\alpha=0$) a clump forms with large number of particles in a small area. This clump is moving and gets reflected from the walls while at some times it is divided to smaller clumps due to the noise and at other times small clumps join to make a bigger clump. \subref{snapshots:negative_alpha} Setting $\alpha=-0.01$ clusters of particles continuously divide and join during simulation. \subref{snapshots:strong_negative_alpha} By further decrease in $\alpha$ vibrating rings emerge.}
\label{fig:snapshots-alpha}
\end{figure}

Moving clumps, moving clusters and vibrating rings could be observed by changing the ratio of repulsion to alignment (Fig.~\ref{fig:snapshots-alpha}). For this purpose we set $\epsilon=0.1$ and change $\alpha$. If we have no alignment ($\alpha = 1$), we obtain a gaseous homogeneous state again, but particles in this homogeneous state have a more robust ballistic motion in comparison to the homogeneous state observed in high noise. A slight decrease in $\alpha$ could lead to formation of a vortex, e.g. Fig. \ref{fig:snapshots-alpha}\subref{snapshots:low_alpha} shows the rotation for $\alpha = 0.01$. Since the repulsion between particles is not strong, they form dense bunches near the walls with a large empty space in the center in such a way that almost all the particles walk on the walls. Increasing the repulsion, more space is covered by the particles. By turning off the repulsion completely, particles form very high density clumps that bounce off the walls [Fig.~\ref{fig:snapshots-alpha}\subref{snapshots:zero_alpha}]. Clumps formation is not observed in periodic boundary condition, because the particles do not meet each other as frequently as confined particles. The shape of clumps in a squared and circular box is square and circle respectively. For negative $\alpha$'s, particles rotate toward each other, and if this attraction is small we see multiple clusters traveling and bouncing off the walls [Fig.~\ref{fig:snapshots-alpha}\subref{snapshots:negative_alpha}]. These clusters are extremely packed and unstable. They may divide into smaller groups or join together to make a bigger mass of particles. The division takes place when particles are more distant than their interaction range so that they cannot return to each other. Finally a strong attraction produces vibrating rings of particles when particles can not escape from the ring [Fig.~\ref{fig:snapshots-alpha}\subref{snapshots:strong_negative_alpha}].

\begin{figure}
\centering
	\subfigure[]{\label{velocity-time-autocorrelation:low-noise}\includegraphics[width=0.48\columnwidth]{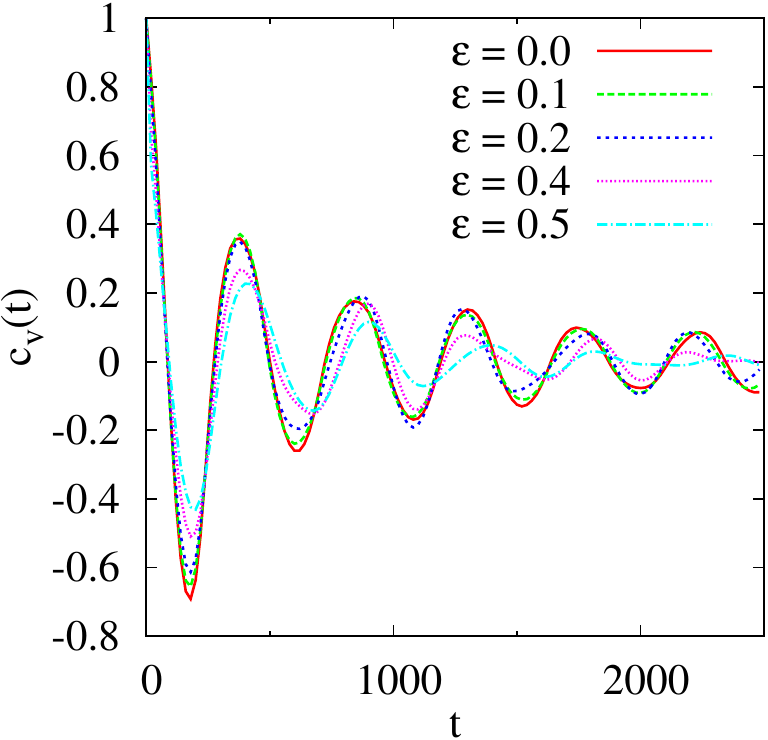}}
	\subfigure[]{\label{velocity-time-autocorrelation:high-noise}\includegraphics[width=0.48\columnwidth]{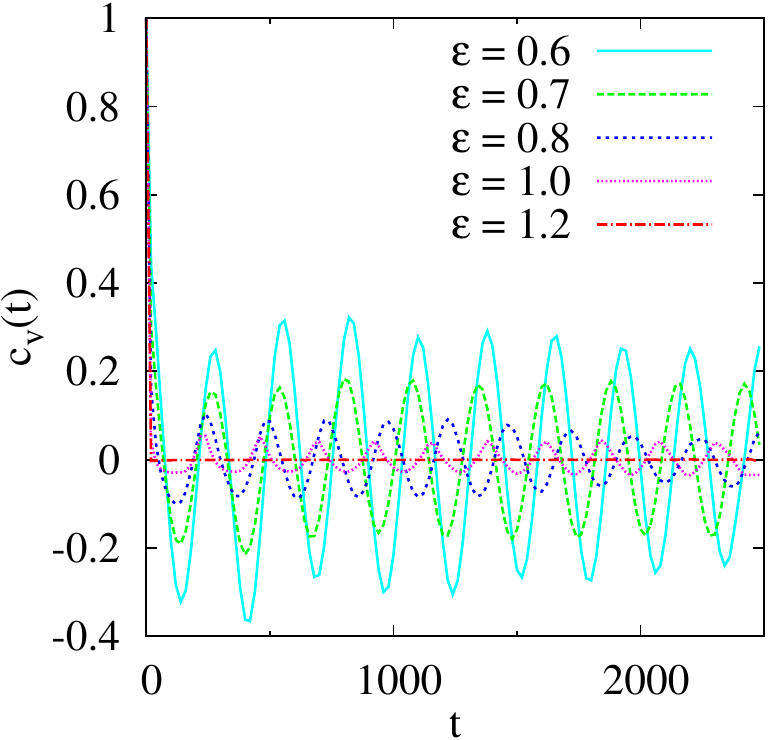}}
	\caption{(Color online) Velocity autocorrelation over time for systems with $\rho_0=1$, $\alpha=0.5$, and various $\epsilon$. \subref{velocity-time-autocorrelation:low-noise} The noise is lower than critical noise and the system is in the vortex phase. \subref{velocity-time-autocorrelation:high-noise} A big polar structure rotating ($\epsilon=0.6$), locally propagating band structures ($\epsilon=0.7,0.8,1.0$), and homogeneous gaseous phase ($\epsilon=1.2$).}
\label{fig:velocity-time-autocorrelation}
\end{figure}

To find detailed properties of these phases we look at velocity autocorrelations over time ($C_v(\tau) = \langle \vec{v}_i(t) \cdot \vec{v}_i(t+\tau) \rangle_{i,t}$), and space ($C_v(r)$). We see that autocorrelations in time oscillate in both vortex phase and band structures (Fig.~\ref{fig:velocity-time-autocorrelation}). The oscillation in vortex phase [Fig.~\ref{fig:velocity-time-autocorrelation}\subref{velocity-time-autocorrelation:low-noise}] is damping over time gradually. This damp is due to the different frequency of rotation of particles. For instance in a vortex in a circular geometry, $C_v(\tau)$ is computed by integrating velocity autocorrelation of individual particles: $C_v(\tau) = 1 / N \int_{r_{min}}^{r_{max}} 2 \pi r \rho(r) \cos(v \tau/r) dr$, where $r_{min}$ and $r_{max}$ are minimum and maximum radii of rotation respectively and $v$ is the speed of particles. One can see that in large $\tau$, the cosine in the integral changes fast and if $\rho(r)$ is a smooth and slow changing function, positive and negative parts of the integral in one oscillation cancel each other. Hence, the integration gives us a negligible result when $\tau$ is large. Instead, in the traveling band structure phase, $C_v(\tau)$ shows more robust oscillations [Fig.~\ref{fig:velocity-time-autocorrelation}\subref{velocity-time-autocorrelation:high-noise}]. Band structures travel in the medium with a constant speed which depends on noise strength. These waves bounce from the walls and periodically move from one corner to another. In homogeneous gaseous state, $C_v(\tau)$ spontaneously reaches zero [Fig.~\ref{fig:velocity-time-autocorrelation}\subref{velocity-time-autocorrelation:high-noise} $\epsilon = 1.2$].

\begin{figure}
\centering
	\subfigure[]{\label{velocity-spatial-autocorrelation:low-noise}\includegraphics[width=0.48\columnwidth]{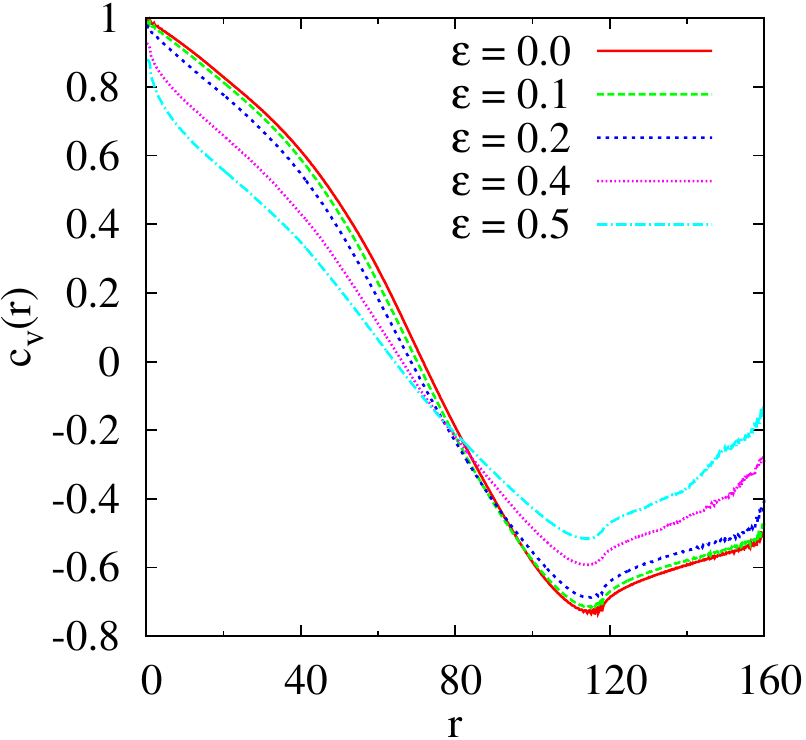}}
	\subfigure[]{\label{velocity-spatial-autocorrelation:high-noise}\includegraphics[width=0.48\columnwidth]{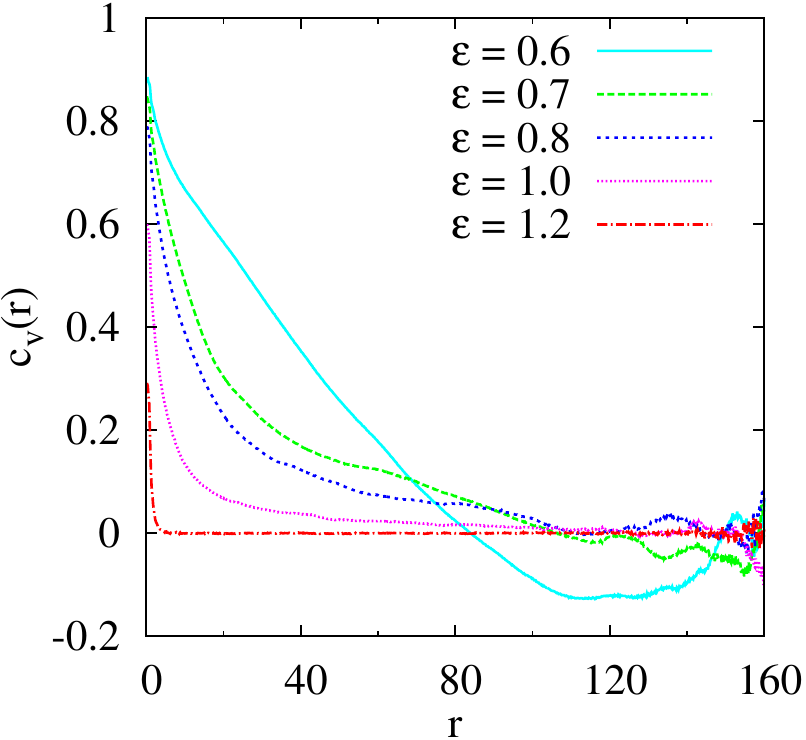}}
	\caption{(Color online) Velocity autocorrelation over distance for systems with $\rho_0=1$, $\alpha=0.5$, and various $\epsilon$. \subref{velocity-spatial-autocorrelation:low-noise} The noise is lower than critical noise and the system is in the vortex phase. \subref{velocity-spatial-autocorrelation:high-noise} A big polar structure rotating ($\epsilon=0.6$), locally propagating band structures ($\epsilon=0.7,0.8,1.0$), and homogeneous gaseous phase ($\epsilon=1.2$).}
\label{fig:velocity-spatial-autocorrelation}
\end{figure}

In vortex phase $C_v(r)$ (Fig.~\ref{fig:velocity-spatial-autocorrelation}) shows the behavior and characteristic length scale of rotation [Fig.~\ref{fig:velocity-spatial-autocorrelation}\subref{velocity-time-autocorrelation:low-noise}]. This length scale is the same as the box dimension. In higher noise no negative correlation in long range distances is seen except for rotating pattern in $\epsilon=0.6$ [Fig.~\ref{fig:velocity-spatial-autocorrelation}\subref{velocity-time-autocorrelation:high-noise}].

\begin{figure}
\centering
\includegraphics[width=\columnwidth,clip]{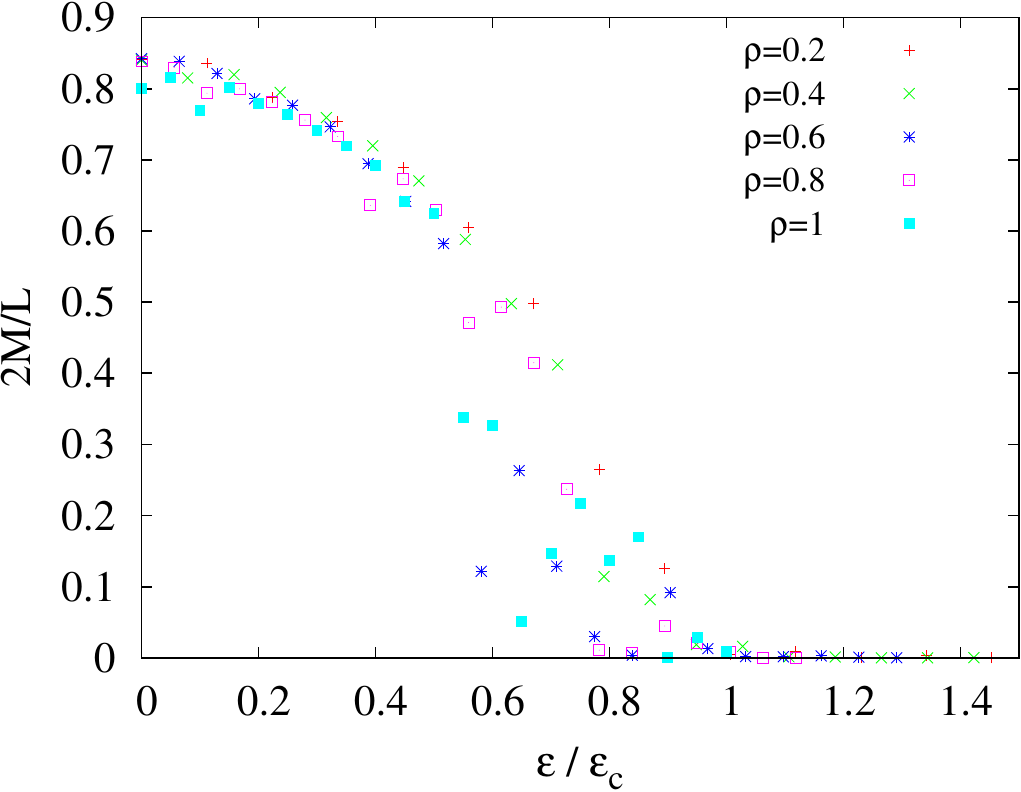}
\caption{(Color online) Scaled angular momentum per particle ($2M/L$) versus scaled noise $\epsilon/\epsilon_c$ for different values of density $\rho_0$. Simulation box size is $L=120$. By decreasing noise, system shows a phase transition. Moreover, data for different densities collapse on the same curve.}
\label{fig:M-noise}
\end{figure}

We claim that the vortex formation is a consequence of the confinement of polar state. This claim is supported by the coincidence of the vortex transition point of confined particles with the non-polar to polar transition point of unconfined particles. To characterize vortex transition we use the average angular momentum of the particles,
\begin{equation}
M = \frac{1}{N} \left\lvert \sum_i^N v \vec{R}_i \times \hat{e}_{\theta_i} \right\rvert.
\end{equation}
Since $M$ scales with $L$, we plot $2M/L$ as a function of scaled noise $\epsilon/\epsilon_c$ in Fig.~\ref{fig:M-noise}. We observe a transition in angular momentum at $\epsilon=\epsilon_c$ for all densities.

\begin{figure}
\centering
\includegraphics[width=\columnwidth,clip]{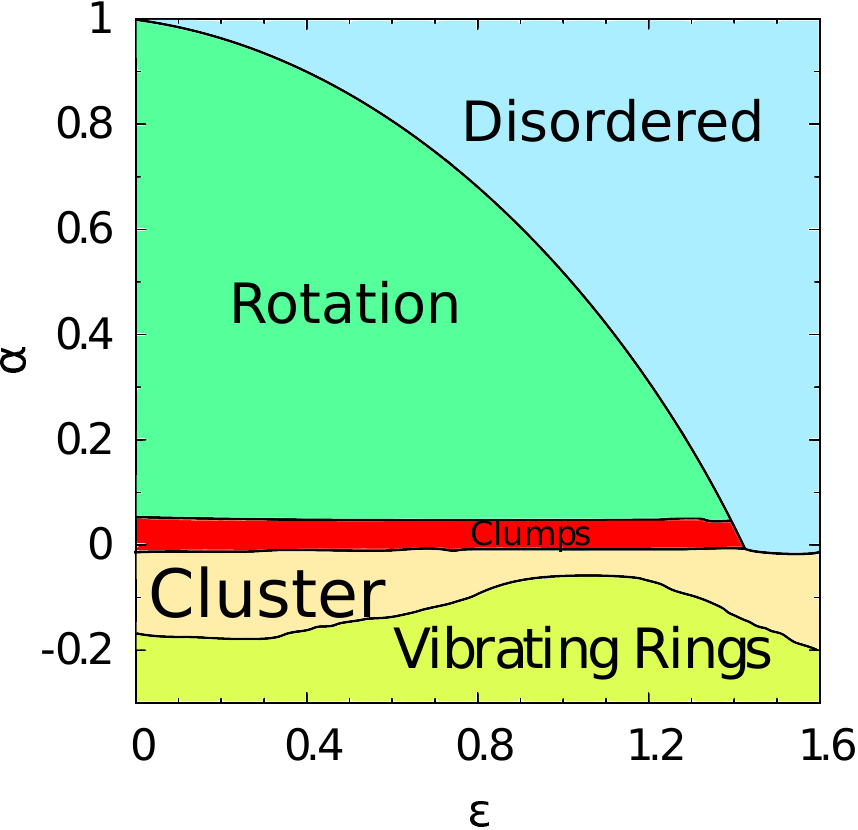}
\caption{(Color online) Phase diagram of the system with $\rho_0=1$ and $L=60$. This diagram is constructed from the results of 625 points, each averaged over 15 realizations. All boundaries are constructed by looking at angular momentum and local cohesion except the boundary between rotation and disordered phase which is from theoretical prediction. Rotation is recognized by looking at angular momentum per particle greater than $0.5$. Vibrating rings where recognized by local cohesion smaller than $0.1$ when $\alpha < 0$. The remaining points are either in the moving clusters or the clumps phase. All points with $\alpha=0$ correspond to formation of moving clumps and the rest of points are in moving clusters regime.}
\label{fig:phase-diagram}
\end{figure}

The phase diagram of the system with six different phases is shown in Fig.~\ref{fig:phase-diagram}. To define vortex phase we looked at angular momentum per particle ($\ave{M} > 0.5$). To recognize vibrating rings and disordered phases, we computed local cohesion,
\begin{equation}
\Phi = \frac{1}{N_w} \left\lvert \sum_{ij} \cos(\theta_i - \theta_j) \right\rvert.
\end{equation}
Here $N_w$ is the number of particles in a certain window and the sum is over particles within that window.

\subsection{Suppressed Spreading and Defect Lines}

\begin{figure}
	\centering
	\subfigure[]{\label{fields:density-low-noise}\includegraphics[height=0.46\columnwidth]{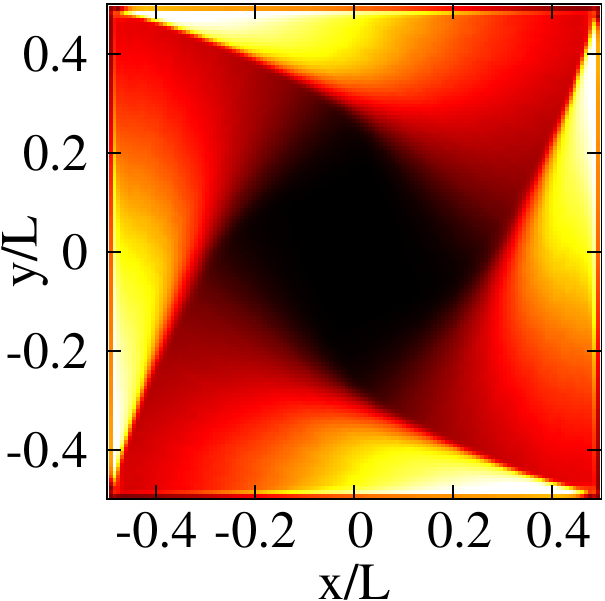}}
	\subfigure[]{\label{fields:density-high-noise}\includegraphics[height=0.46\columnwidth]{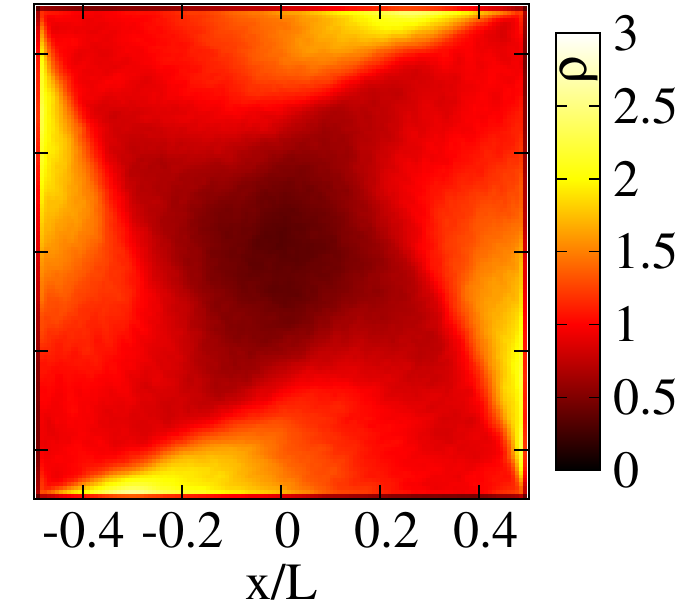}}
	\subfigure[]{\label{fields:cohesion-low-noise}\includegraphics[height=0.46\columnwidth]{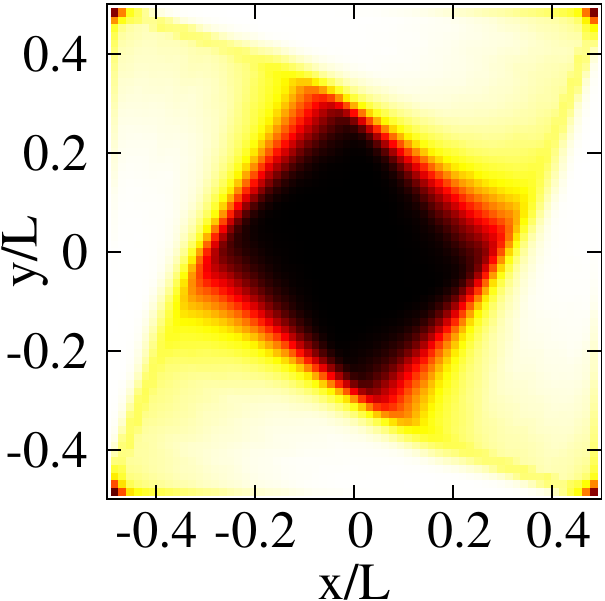}}
	\subfigure[]{\label{fields:cohesion-high-noise}\includegraphics[height=0.46\columnwidth]{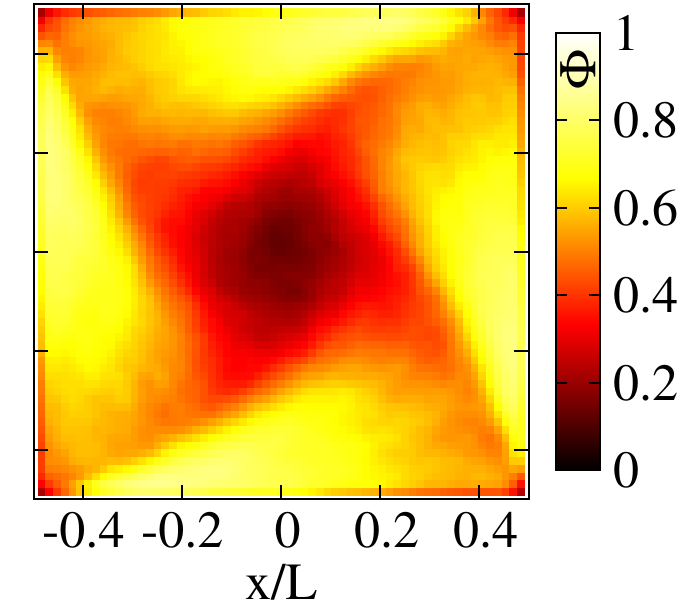}}
	\subfigure[]{\label{fields:velocity-low-noise}\includegraphics[height=0.46\columnwidth]{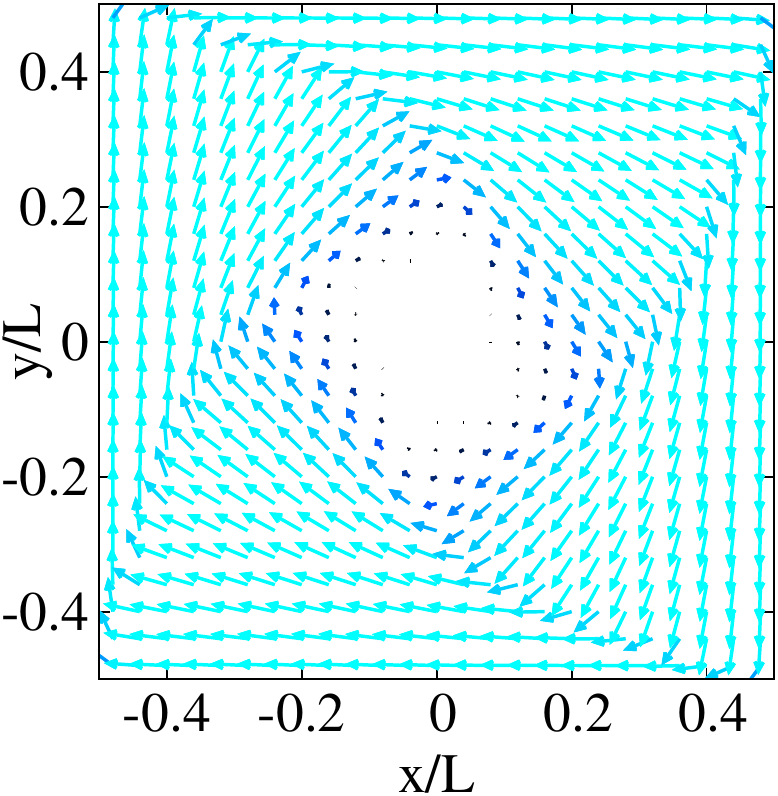}}
	\subfigure[]{\label{fields:velocity-high-noise}\includegraphics[height=0.46\columnwidth]{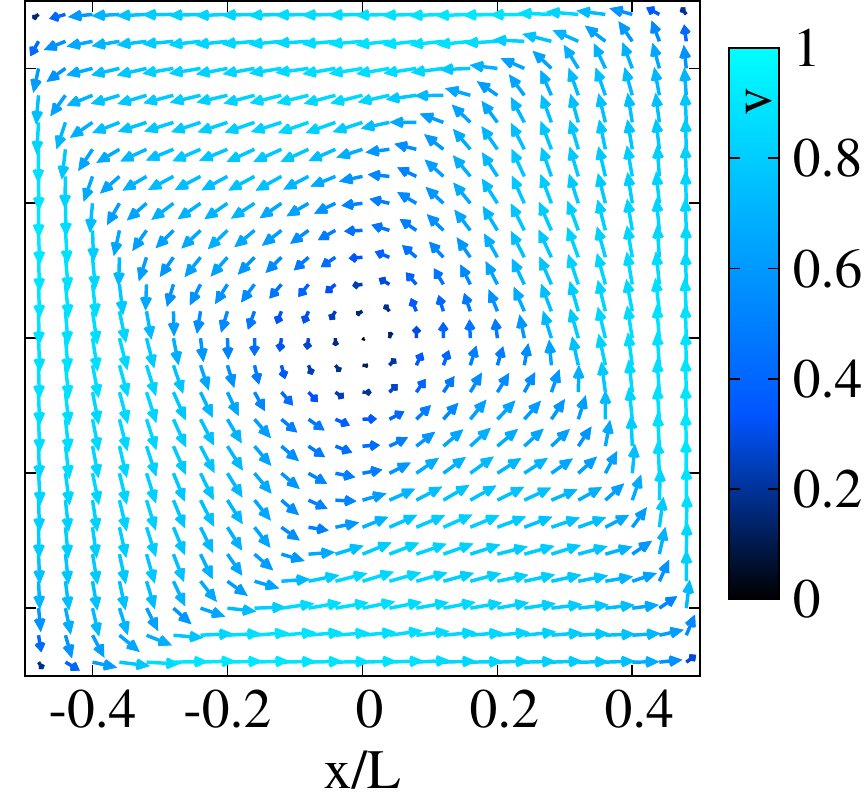}}
	\caption{(Color online) Time average of $\rho$, $\Phi$ and $\vec{v}$ in space for $\rho_0 = 1$ and $\epsilon = 0,0.5$ in a box of size $L=120$. \subref{fields:density-low-noise} and \subref{fields:density-high-noise} density with grid dimension 128 by 128. \subref{fields:cohesion-low-noise} and \subref{fields:cohesion-high-noise} cohesion with grid dimension 64 by 64. \subref{fields:velocity-low-noise} and \subref{fields:velocity-high-noise} velocity with grid dimension 32 by 32. Color codes correspond to the magnitude of each field and the length of arrows in \subref{fields:velocity-low-noise} and \subref{fields:velocity-high-noise} shows the average speed. \subref{fields:density-low-noise},\subref{fields:cohesion-low-noise}, and \subref{fields:velocity-low-noise} belong to the same noiseless simulation of a clockwise vortex ($\epsilon=0$). \subref{fields:density-high-noise}, \subref{fields:cohesion-high-noise}, and \subref{fields:velocity-high-noise} belong to the simulation of a counter clockwise vortex with noise ($\epsilon=0.5$). Defect lines corresponding to suppressed spreading behavior are observed in reagion with low cohesion and high density.}
\label{fig:fields}
\end{figure}

\begin{figure}
	\centering
	\subfigure[particle simulations]{\label{fig:compare-md-density}\includegraphics[height=0.46\columnwidth]{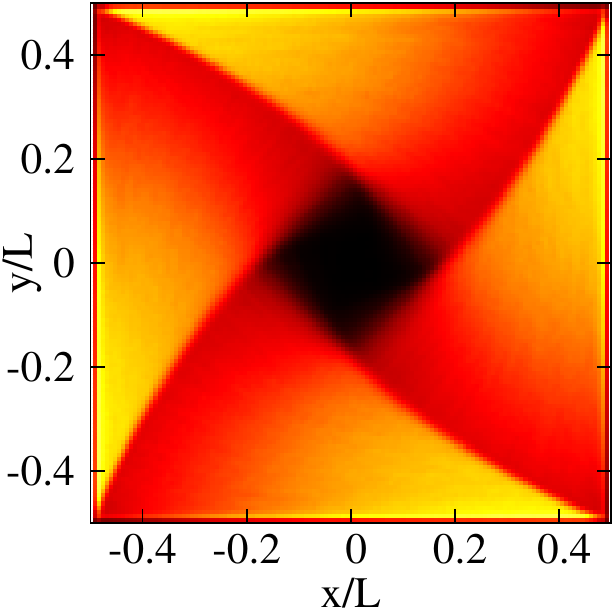}}
	\subfigure[continuum model]{\label{fig:compare-continuum-density}\includegraphics[height=0.46\columnwidth]{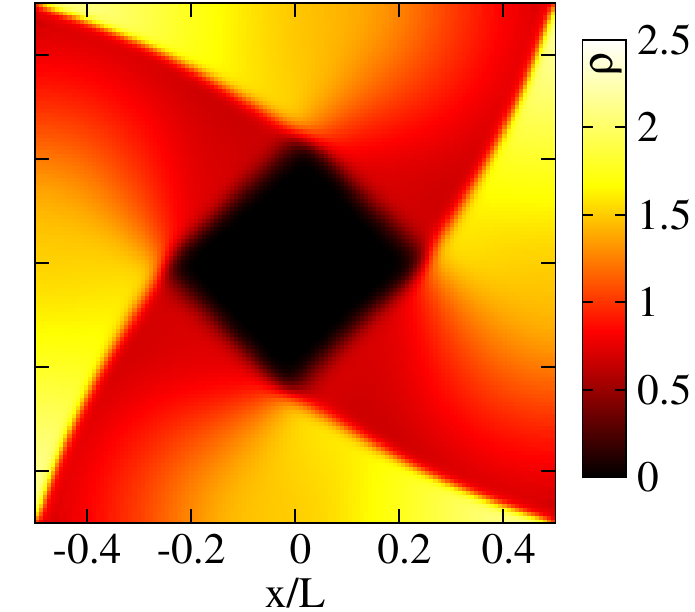}}
	\caption{(Color online) Comparison of average density field for \subref{fig:compare-md-density} particle model and \subref{fig:compare-continuum-density} continuum model. Parameters are set to $\rho_0 = 1$, $g = 2$, $\alpha = 0.5$, $\epsilon = 0$, and $L = 120$. The color represents density. The grid size is 128 by 128. Few points of continuum model are in range $\rho \in (-0.1,0)$ but to have a better comparison we plot both maps in the same interval $\rho \in (0,2.5)$.}
\label{fig:compare-density}
\end{figure}

Suppressed spreading and presence of the defect lines are two of the interesting features of the system. For a clearer observation of defect lines we could measure the cohesion between particles. Figure \ref{fig:fields} shows time averaged $\rho(\vec{r})$, $\Phi(\vec{r})$, and $\vec{v}(\vec{r})$. Near the corners, density gets its highest value. Velocity field shows an outgoing flow of particles at each corner which is suppressed by collision of incoming flow to the same corner. This collision makes a defect line that corresponds to a lower cohesion but high density. The name defect line is used, due to the spontaneous change of velocity in the boundary between incoming and outgoing flows.

\begin{figure}
	\centering
	\subfigure[particle simulations]{\label{fig:compare-md-velocity}\includegraphics[height=0.46\columnwidth]{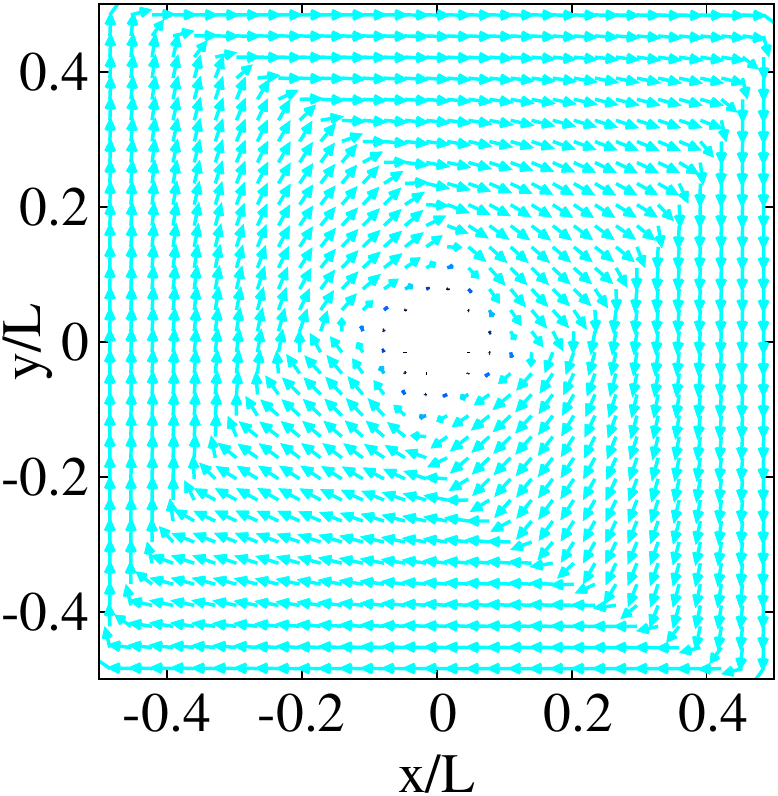}}
	\subfigure[continuum model]{\label{fig:compare-continuum-velocity}\includegraphics[height=0.46\columnwidth]{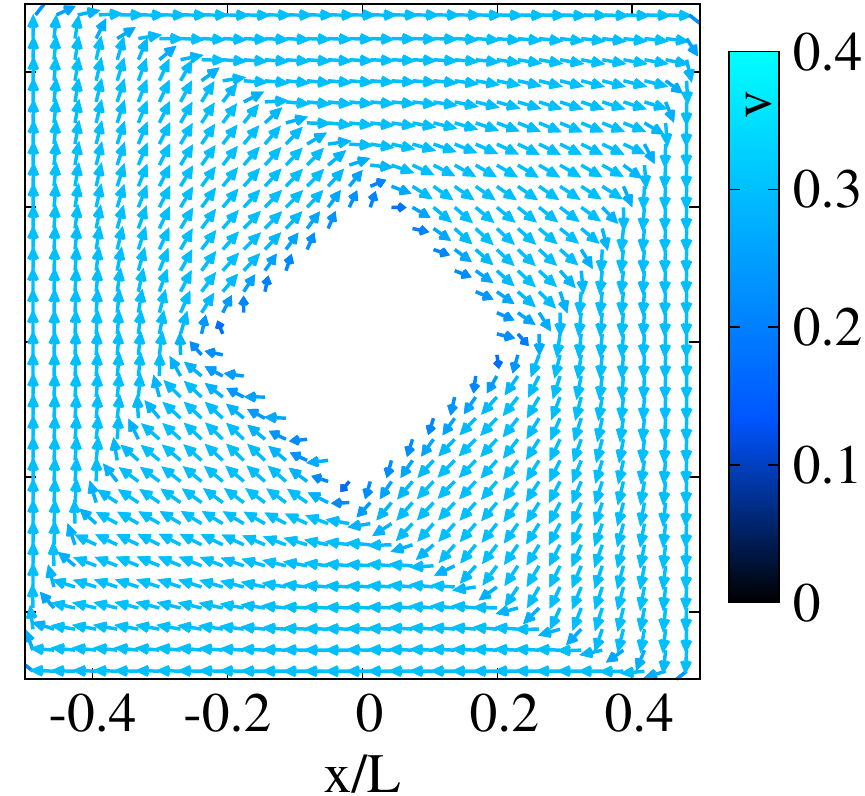}}
\caption{(Color online) Comparison of average velocity field for \subref{fig:compare-md-velocity} particle model and \subref{fig:compare-continuum-velocity} continuum model. The color represents velocity magnitude. The length of arrows is proportional to the speed. The grid size is 32 by 32. To enhance the velocity map of the continuum model, we removed points with density lower than $0.03$ that have large computational errors.}
\label{fig:compare-velocity}
\end{figure}

Defect lines are also present in the solution of the continuum model introduced in subsection \ref{subsection:numerical-method}. Density and velocity fields in the numerical solution and particle simulations are plotted in Figs.~\ref{fig:compare-density} and \ref{fig:compare-velocity} respectively, for a system with $\rho_0=1$, $g_p=2$, $\alpha=0.5$ and $v=0.3$ which show excellent qualitative but not precise quantitative agreement between the simulation and the continuum model.

\begin{figure}
	\centering
	\subfigure[$\rho,\Phi$ along $y_0 = 53$]{\label{field-section-y-53:density}\includegraphics[width=0.49\columnwidth]{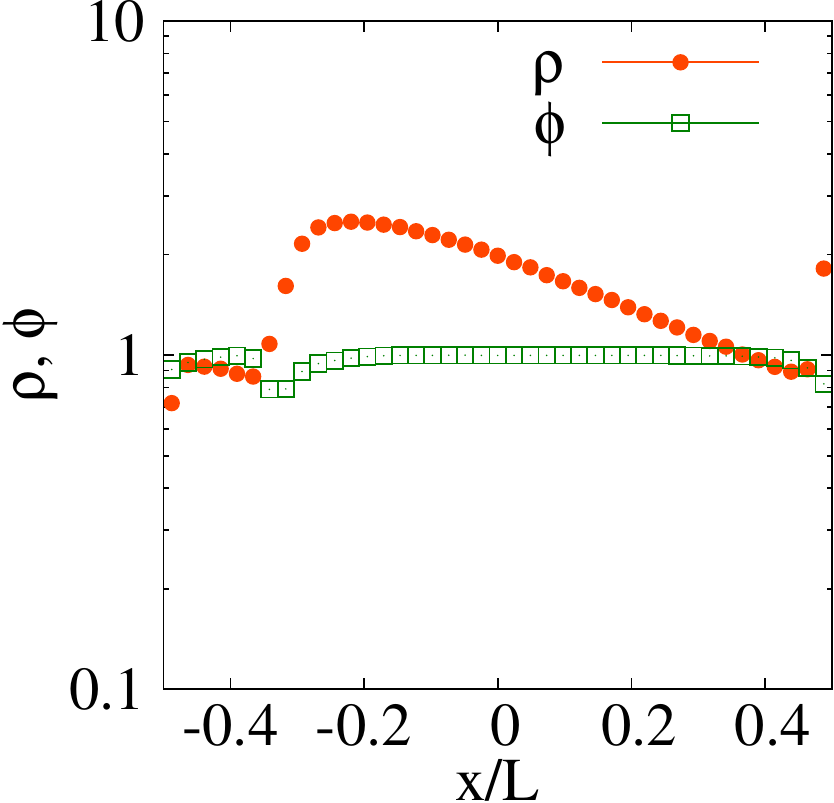}}
	\subfigure[$\vec{v}$ along $y_0 = 53$]{\label{field-section-y-53:velocity}\includegraphics[width=0.49\columnwidth]{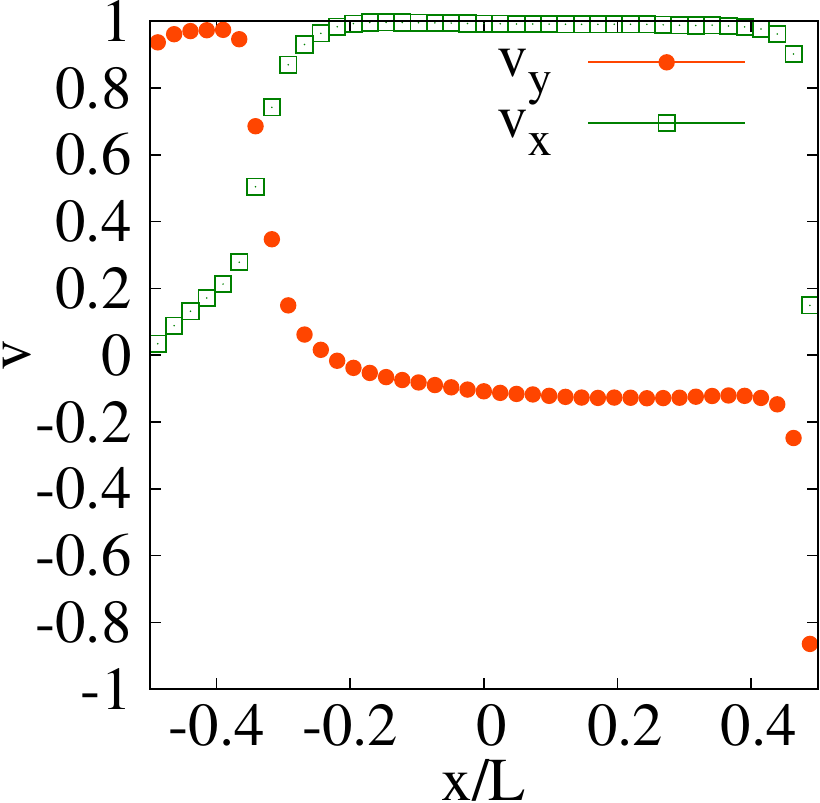}}
	\subfigure[$\omega$ on line $y_0 = 0$]{\label{field-section-y-0:omega}\includegraphics[width=0.49\columnwidth]{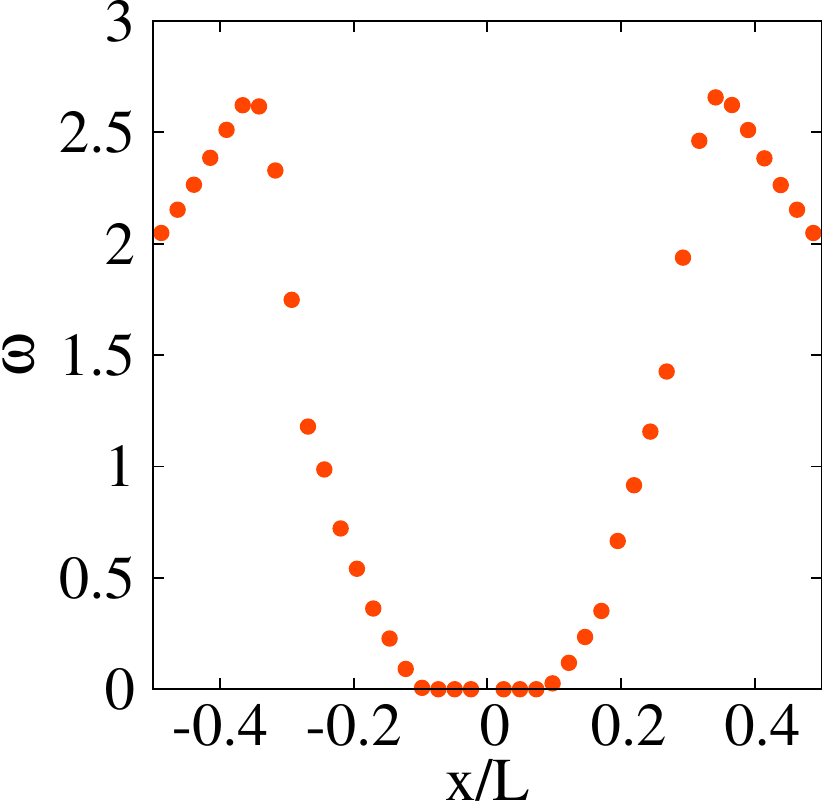}}
	\subfigure[$\vec{v}$ on line $y_0 = 0$]{\label{field-section-y-0:velocity}\includegraphics[width=0.49\columnwidth]{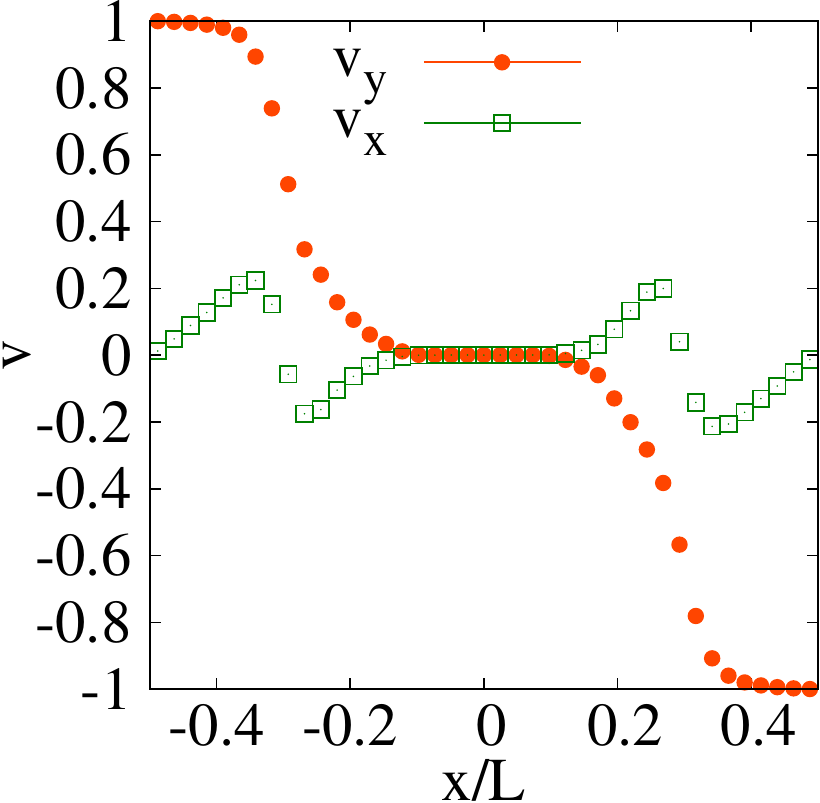}}
\caption{(Color online) Cross sectional value of time averaged fields. \subref{field-section-y-53:density} $\rho$ and $\Phi$ across horizontal line $y=53$. \subref{field-section-y-53:velocity} and \subref{field-section-y-0:velocity} $v_x$ and $v_y$ across horizontal lines \subref{field-section-y-53:velocity} $y=53$, and \subref{field-section-y-0:velocity} $y=0$. \subref{field-section-y-0:omega} Angular velocity $\omega$ across horizontal line $y=0$. Simulation parameters are $\rho_0 = 1.0$, $\epsilon=0.05$ and $L=120$. Sub-caption of each plot shows the value of $y_0$ and the corresponding value. One can see that density exponentially decreases and the cohesion has a fall-off corresponding to crossing the defect line in \subref{field-section-y-53:density}. Velocity component perpendicular to the wall is constant along the wall \subref{field-section-y-53:velocity} and it linearly increases with distance from the wall \subref{field-section-y-0:velocity}. Finally \subref{field-section-y-0:omega} shows that rotation frequency of particles depends on their position.}
\label{fig:field-section-y-changing}
\end{figure}

Figure \ref{fig:field-section-y-changing} shows the change in average density, cohesion, velocity and rotational frequency of particles across a horizontal line $y=y_0$ for $y_0 = 53$ and $y_0 = 0$. One can see that right after defect (lower cohesion), density is exponentially decreasing [Fig.~\ref{fig:field-section-y-changing}\subref{field-section-y-53:density}]. By looking at the velocities we find that velocities are tangent to the wall and by getting away from the wall we see that perpendicular component of velocity to the wall is independent of horizontal position [Fig.~\ref{fig:field-section-y-changing}\subref{field-section-y-53:velocity}], and is proportional to the distance from the wall [Fig. \ref{fig:field-section-y-changing}\subref{field-section-y-0:velocity}]. This proportionality corresponds to the exponential shape of the defect line. To show this, we first define $u(x)$ the distance between the top defect line in Fig.~\ref{fig:fields}\subref{fields:cohesion-low-noise} from the top wall as a function of $x$. Linear relation of perpendicular component of velocity to the wall gives us $v_y = a u$ where $a$ is a positive number. This linear relation lets us find a differential equation with exponential solution:
\begin{equation}
\frac{d u(x)}{d x} = - \frac{v_y}{v_x} = - \frac{v_y}{\sqrt{1 - v^2_y}} \approx -v_y = a u(x).
\label{eq:dudx}
\end{equation}
Here we used the approximation that $\left| v_y \right| \ll 1$ which is clear from Fig.~\ref{fig:field-section-y-changing}\subref{field-section-y-0:velocity}.

\begin{figure}
\centering
\includegraphics[width=\columnwidth,clip]{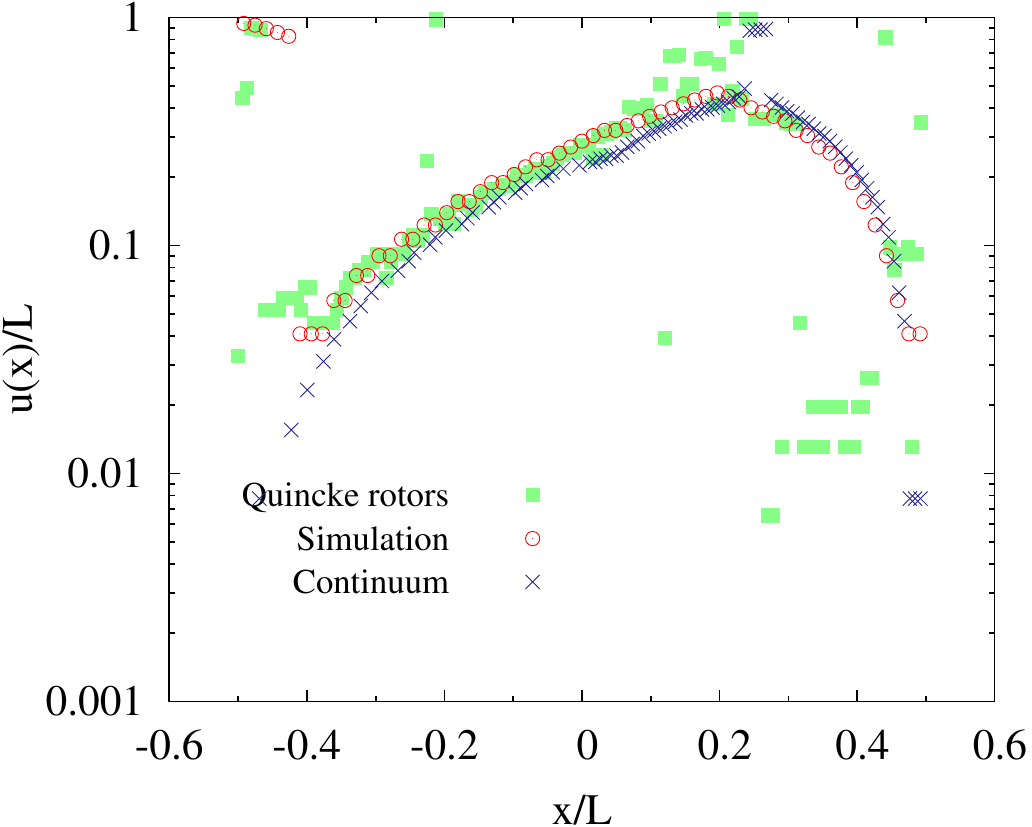}
\caption{(Color online) The defect line distance from the wall $u$ as a function of horizontal position $x$ from the corner in three cases of Quincke rotor experiment, simulation and continuum theory. Simulation and continuum parameters are set to $\rho_0=1$, $\alpha=0.5$, $v=0.3$,$\epsilon=0$, and $L=120$. Experimental data are averaged movie frames of reference \cite{Bricard2013} with enhancement of final image to avoid computational error and whiteness considered proportional to density. $u$ is defined as the points where $\left| \frac{1}{\rho} \Dif{\rho}{y} \right|_{(x,L/2 - u(x))} > 0.1$. One can see that at the interval $-0.3 < x/L < 0.2$, the defect line shape is exponential which is in agreement with the exponential decay of density in Fig.~\ref{fig:field-section-y-changing}\subref{field-section-y-53:density}. We also see that experiment, simulation and continuum model are all in good agreement.}
\label{fig:boundary}
\end{figure}

Defect lines in the experiment of reference \cite{Bricard2013} also seem to have exponential form. We extracted experimental images from a movie in supplementary material of Quincke rotor experiment \cite{Bricard2013} and assumed the averaged gray scale over the images is proportional to the density. The form of defect lines in simulation, continuum model, and experiment are sketched in Fig.~\ref{fig:boundary} that shows a very good agreement between these three data sets.

Figure \ref{fig:field-section-y-changing}\subref{field-section-y-0:omega} shows that the angular velocity of particles across $y=0$ is not constant. We also computed the average angular velocity ($\omega = \Delta \theta / \Delta t$) of any particle around the center of the box and observed that the angular velocity is almost proportional to the inverse of distance from the center ($\omega \sim 1/r$). That means particles velocity component perpendicular to their position vector is rather constant ($\omega = v_{\perp} / r$). The result shows that in the low noise regime $v_{\perp} \approx \ave{v} \approx v = 1$; However, a reduction in $v_{\perp}$  as well as the time averaged velocity of the particles is observed when the noise is high.

\section{Discussion}

In this study we presented a minimal model to mimic the behavior of Quincke rotors in a squared box. Our model has alignment and repulsion between particles and it shows six different phases that are homogeneous disordered, moving clumps, moving clusters, vibrating rings, traveling stripes and vortex. Our focus in this paper was to study the vortex formation. We derived hydrodynamic equations in high noise and low noise limits and saw that each of them could describe some aspect of the system. With high noise equations we are able to predict the transition point while we are not able to find a vortex formation; However, low noise equations give us vortex solutions. The shape of vortex resulting from continuum model is very similar to our simulations and the Quincke rotors experiment. In the theoretical model, simulations and experiment, the center of box is rather empty and we observe a four-fold symmetry in the shape of the vortex. This four-fold symmetry is because the spread of particle flows going out of a corner is suppressed by collision of another flow and a defect line emerges for each corner. The shape of this defect line is exponential in experiment, theory and simulations and we brought some evidences to prove it. In addition to the shape of defect line, our theoretical calculation well predicts the size of empty region in the center of the box and shows that the size depends on repulsion, total density of particles and their velocity. The velocity dependence is important and one could use self-propelled particles inside a box, as separator of fast and slow moving particles. Because Quincke rotors velocities (Eq. \ref{eq:Quicke-rotor-velocity}) depend on their radius and other environmental factors~\cite{Bricard2013}, they are particular candidate for future experiment about this separation technique. Our next aim of research is to study the behavior of mixture of fast and slow particles, using simulations. Other studies of similar systems could be done like a more precise study of transition with finite size scaling or a study of other geometries of the boundary, e.g. $n$-sided polygon. One could also find the limit of $n$ in an $n$-sided polygon in which the vortex does not have the symmetry of the polygon anymore. Finally, one can construct hydrodynamic equations using Gaussian approximation and check the accuracy of the results~\cite{sonnenschein2013approximate}.

\begin{acknowledgments}
We are grateful to Julien Tailleur for a stimulating conversation, Lutz Schimansky-Geier, Stefano Ruffo, Fernando Peruani, Siriam Ramaswamy and Igore Aronson for their comments on the project. We also thank Maryam Khatami, Sayeh Rajabi and Bernard Sonnenschein for critical reading of the manuscript. This work was supported by Iran national science foundation (93031724). We also thank GGI and the Humboldt University of Berlin for their hospitality and the Humbold University of Berlin for providing computational resources and financial support (IRTG 1740).
\end{acknowledgments}

\bibliography{references}

\end{document}